\documentclass[a4paper,fleqn,usenatbib]{mnras}

\pdfoutput=1

\usepackage{newtxtext,newtxmath}
\usepackage{multirow}
\usepackage{tablefootnote}
\usepackage{threeparttable}

\usepackage[T1]{fontenc}
\usepackage{ae,aecompl}

\usepackage{graphicx}	
\usepackage{epstopdf}
\usepackage{amsmath}	
\usepackage{amssymb}	
\usepackage{subcaption}
\usepackage{braket}
\usepackage[]{hyperref}

\title[CO kinematics in local radio galaxies]{\centering{The AGN fuelling/feedback cycle in nearby radio galaxies \\
  II. Kinematics of the molecular gas}}

\author[I. Ruffa et al.]{Ilaria Ruffa,$^{1,2}$\thanks{E-mail: i.ruffa@ira.inaf.it}
Timothy A. Davis,$^{3}$
Isabella Prandoni,$^{2}$
Robert A. Laing,$^{4}$
\newauthor
Rosita Paladino,$^{2}$
Paola Parma,$^{2}$
Hans de Ruiter,$^{2}$
Viviana Casasola,$^{2}$
\newauthor
Martin Bureau,$^{5,6}$
and Joshua Warren$^{5}$
\\
$^{1}$Dipartimento di Fisica e Astronomia, Universit\`{a} degli Studi di Bologna, via P.\ Gobetti 93/2, 40129 Bologna, Italy\\
$^{2}$INAF - Istituto di Radioastronomia, via P.\ Gobetti 101, 40129 Bologna, Italy\\
$^{3}$School of Physics \& Astronomy, Cardiff University, Queens Buildings, The Parade, Cardiff CF24 3AA, UK\\
$^{4}$Square Kilometre Array Organisation, Jodrell Bank Observatory, Lower Withington, Macclesfield, Cheshire SK11 9DL, UK\\
$^{5}$Sub-dept.\ of Astrophysics, Dept.\ of Physics, University of Oxford, Denys Wilkinson Building, Keble Road, Oxford OX1 3RH, UK\\
$^{6}$Yonsei Frontier Lab and Department of Astronomy, Yonsei University, 50 Yonsei-ro, Seodaemun-gu, Seoul 03722, Republic of Korea
}

\date{Accepted XXX. Received YYY; in original form ZZZ}

\pubyear{2019}

\def\co{ $^{12}$CO(2-1)}
\def\cotoh2{CO-to-H$_{2}$}

\begin{document}
\label{firstpage}
\pagerange{\pageref{firstpage}--\pageref{lastpage}}
\maketitle

\begin{abstract}
This is the second paper of a series exploring the multi-component (stars, warm and cold gas and radio jets) properties of a sample of eleven nearby low excitation radio galaxies (LERGs), with the aim of better understanding the AGN fuelling/feedback cycle in these objects. Here we present a study of the molecular gas kinematics of six sample galaxies detected in $^{12}$CO(2-1) with ALMA. In all cases, our modelling suggests that the bulk of the gas in the observed (sub-)kpc CO discs is in ordered rotation. Nevertheless, low-level distortions are ubiquitous, indicating that the molecular gas is not fully relaxed into the host galaxy potential. The majority of the discs, however, are only marginally resolved, preventing us from drawing strong conclusions. NGC\,3557 and NGC\,3100 are special cases. The features observed in the CO velocity curve of NGC\,3557 allow us to estimate a super-massive black hole (SMBH) mass of $(7.10\pm0.02)\times10^{8}$~M$_{\odot}$, in agreement with expectations from the M$_{\rm SMBH}- \sigma_{*}$ relation. The rotation pattern of NGC\,3100 shows distortions that appear to be consistent with the presence of both a position angle and inclination warp. Non-negligible radial motions are also found in the plane of the CO disc, likely consistent with streaming motions associated with the spiral pattern found in the inner regions of the disc. The dominant radial motions are likely to be inflows, supporting a scenario in which the cold gas is contributing to the fuelling of the AGN.
\end{abstract}




\section{Introduction}\label{sec:intro}
Active galactic nuclei (AGN) are associated with the accretion of material onto super-massive black holes (SMBHs), located at the centres of their host galaxies. It is now widely believed that AGN activity plays an important role in shaping galaxies over cosmic time, with AGN feedback being responsible of changing the physical conditions of the surrounding inter-stellar medium (ISM) or expelling it from the nuclear regions, thus impacting the star formation processes and the subsequent evolution of the host galaxy \citep[e.g.][]{Combes17,Garcia14,Harrison17}.

In the local Universe, two main AGN modes are commonly discussed, differentiated by the nature of the dominant energetic output \citep[i.e. feedback; for a review see e.g.][]{Fabian12}. The first is the so-called \textit{radiative} mode (also known as \textit{quasar} or \textit{wind} mode), whereby the accretion occurs at high rates ($\ga 0.01~\dot{M}_{\rm Edd}$, where $\dot{M}_{\rm Edd}$ is the Eddington accretion rate\footnote{$\dot{M}_{\rm Edd} = \dfrac{4\pi \, G \, M_{\rm SMBH} \, m_{\rm p}}{\varepsilon \, c \, \sigma_{\rm T}}$, where G is the gravitational constant, $M_{\rm SMBH}$ is the mass of the central super-massive black hole, $m_{\rm p}$ is the mass of the proton, $\varepsilon$ is the accretion efficiency, $c$ is the speed of light and $\sigma_{\rm T}$ is the cross-section for Thomson scattering.}) through an optically thick and geometrically thin accretion disc \citep{Shakura73}. The accretion process in this type of AGN is radiatively efficient, so the dominant energetic output is in the form of electromagnetic radiation produced by the efficient conversion of the potential energy of the gas accreted by the SMBH. The second AGN mode is the \textit{kinetic} mode, whereby the accretion process occurs in a radiatively inefficient way (i.e. little radiated energy), and the dominant energetic output is in kinetic form through the radio jets \citep[e.g.][]{Merloni07}. In this AGN mode material is accreted at low rates ($\ll 0.01 \dot{M}_{\rm Edd}$), with the classic geometrically thin accretion disc either absent or truncated at some inner radius and replaced by geometrically thick, optically thin structures such as advection-dominated accretion flows (ADAFs; \citealp[e.g.][]{Nara95}). Kinetic mode objects are a very intriguing class of AGN: they lack all the observational features typical of "conventional" AGN (e.g.\,there is no clear evidence for the presence of the infrared-emitting dusty torus, accretion-driven emission lines are absent in their optical spectra etc.), indicating they may represent a distinct class of sources, with a distinct formation history.

In the local Universe, radio galaxies (RGs), which by definition show strong kinetic (jet-induced) feedback, can be divided into two main classes according to their optical spectra (e.g.\ \citealt{Heckman14}). High-excitation radio galaxies (HERGs) have spectra showing strong, quasar/Seyfert-like emission lines and are radiatively efficient, thereby producing radiative as well as kinetic feedback. Low-excitation radio galaxies (LERGs) show no or weak, LINER-like emission lines in their optical spectra and their feedback is almost entirely kinetic. Locally, LERGs are the dominant radio galaxy population \citep[e.g.][]{Hardcastle07} and they are typically hosted by very massive early-type galaxies (ETGs), with absolute K-band magnitude $M_{\rm K} \leq -25$~mag (corresponding to $M_{*} \geq 10^{11}~M_{\odot}$; \citealp[e.g.][]{Best12}). Despite their prevalence, however, the trigger mechanisms of these objects and associated AGN feeding/feedback processes are still poorly understood \citep[e.g.][]{Hardcastle18}: investigating the nature of LERGs is thus crucial to shed light on the mechanisms which determine the observed properties of massive ETGs in the local Universe. 

LERGs are generally dominated by an old stellar population, already formed at $z=2$ \citep[e.g.][]{Thomas05,Greene15}. Their red optical colours place them in the so-called "red sequence" in the optical color-magnitude diagram \citep[e.g.][]{Faber07,Salim07}. For this reasons, these objects have traditionally been believed to be mostly devoid of cool ISM reservoirs. This has been one of the strongest argument favouring the hypothesis that LERGs are powered by accretion directly from the hot phase of the inter-galactic medium (IGM; \citealp[e.g.][]{Hardcastle07}) through Bondi spherical accretion \citep{Bondi52}. Although too simplistic in describing the circumnuclear environment of kinetic-mode AGN, the hot accretion scenario was initially supported by the finding of a correlation between the jet power in LERGs and the Bondi accretion rate \citep[][]{Allen06,Balmaverde08,Russell13}.

Over the past decade, however, this "hot-mode" picture has been partly questioned. on the theoretical side, more realistic models for the accretion process in LERGs have been proposed on the basis of different numerical simulations and are now referred to as \textit{chaotic cold accretion} models \citep[e.g.][]{King07,Wada09,Naya12,Gaspari13,King15,Gaspari15,Gaspari17}. In this mechanism the hot gas from the galaxy halo cools because of turbulence to temperatures lower than 10$^{3}$~K prior to the accretion onto the SMBH. 
Hints of cold gas clouds falling towards active nuclei have recently been observed in some objects \citep[e.g.][]{Tremblay16,Maccagni18}, providing support for this hypothesis.
Observationally, cold (atomic and molecular) gas and dust are often detected in LERGs, with masses that are potentially capable of powering the jets by accretion  \citep[$M_{\rm H_2} \sim 10^{7} - 10^{10}$~M$_{\rm \odot}$, e.g. ][]{Okuda05,Prandoni07,Prandoni10,Ocana10,Ruffa19a}.

The presence of cold gas alone is not a direct evidence of fuelling, however. For example, the cold gas in 3C\,31 (the prototypical LERG) is found to be mostly in ordered rotation and stable orbits (\citealt[][]{Okuda05}, North et al., submitted): in cases like this the accretion rate may be relatively low. Furthermore, the few existing spatially-resolved studies of molecular gas in radio-loud objects show that the molecular gas is outflowing or interacting with the radio jets, rather than infalling \citep[e.g.][]{Alatalo11,Combes13,Oosterloo17}. Indeed, jets expanding into the surrounding  material  can  create shells of shocked gas that are pushed away from the jet axis, resulting in lateral expansion (i.e.\,outflow; \citealp{Wagner12}).  Another possible form of interaction could be entrainment of the molecular gas, causing fragmentation of the gas clouds and jets deceleration from relativistic speeds \citep[e.g.][]{Laing02}. In both cases, however, the jet-cold gas mutual interplay leaves clear signs in the gas distribution and kinematics (i.e.\,clumpy/disrupted morphology; kinematical distortions/asymmetries).
Resolved studies of LERGs are currently very sparse: detailed investigation of the physical and kinematical properties of the cold gas in the nuclear regions of these objects would represent a fundamental step-forward in our understanding on the fuelling/feedback mechanism in LERGs.

We aim to carry out an extensive, multi-phase study of the various galaxy components (stars, warm and cold gas, radio jets) to get a better understanding of the AGN fuelling/feedback cycle in LERGs. Our specific project is a systematic study of a complete volume- and flux-limited ($z<0.03$, $S_{2.7 \rm GHz}\leq 0.25$~Jy) sample of eleven LERGs in the southern sky \citep[][hereafter Paper I]{Ruffa19a}. In Paper I, we presented ALMA Cycle 3 $^{12}$CO(J=2-1) and 230~GHz continuum observations of nine objects, at spatial resolutions of few hundreds of parsec. Our work shows that rotating (sub-)kpc molecular discs are very common in LERGs: six out of nine sources observed with ALMA have been detected in CO. 
We present here the 3D modelling of these six CO discs. The paper is structured as follows. In Section~\ref{sec:obs} we briefly summarise the ALMA observations used in this work. The general method used to model the CO(2-1) discs is described in Section~\ref{sec:method}. Details of the modelling of individual sources are reported in Section~\ref{sec:individual_sources}. We discuss the results in Section~\ref{sec:discussion}, before summarising and concluding in Section~\ref{sec:conclusion}. 

Throughout this work we assume a standard $\Lambda$CDM cosmology with H$_{\rm 0}=70$\,km\,s$^{-1}$\,Mpc$^{\rm -1}$, $\Omega_{\rm \Lambda}=0.7$ and $\Omega_{\rm M}=0.3$.

\section{ALMA observations}\label{sec:obs}
Full details of our ALMA observations and data reduction can be found in Paper I; a brief summary is presented here. 

CO(2-1) ALMA observations were taken during Cycle 3, between March and July 2016 (PI: I.\ Prandoni). The CO(2-1) line (rest frequency 230.5380~GHz) was observed using the high-resolution correlator configuration (1920 1.129~MHz-wide channels), providing about three raw channels in a 5~km~s$^{-1}$ bin. The array configurations provide spatial resolutions (at $\approx$230~GHz) ranging from $\approx0.3$ to $\approx0.9$~arcsec, corresponding to few hundreds of parsecs at the redshift of our sources. 
Titan and Pallas were used as primary flux calibrators; bright quasars were observed as standards if no solar-system object was available.

The data were calibrated and imaged using the Common Astronomy Software Application \citep[{\sc casa};][]{McMullin07} package, version 4.7.2. The continuum-subtracted CO data cubes were produced using the \texttt{clean} task with natural weighting (see Paper I for details). The final channel widths range from 10 to 40~km~s$^{-1}$, chosen to achieve the best compromise between the signal-to-noise ratio (S/N) and the sampling of the line profiles. 

\subsection{Archival ALMA Data}\label{sec:archival_data}
Higher-resolution CO(2-1) ALMA data of one of our sample sources (NGC\,3557) were taken during Cycle 4 and are now publicly available in the ALMA archive (project 2015.1.00878.7, PI: Barth). The CO(2-1) disc of NGC\,3557 is barely resolved at the resolution of our Cycle 3 ALMA observations (0.6$''\simeq130$~pc; Paper I).
To better constrain the kinematics of the molecular gas, in this work we combined the two datasets.

The Band 6 archival data of NGC\,3557 were taken in July 2016. The spectral window centred on the redshifted frequency of the CO(2-1) line is composed of 480 3.906~MHz-wide channels. 39 12-m antennas were arranged in an extended configuration, with a maximum baseline length of 1.1~km. The achieved spatial resolution is $\approx0.4$~arcsec ($\approx90$~pc). We calibrated the data manually, using the same approach as for the other observations (see Paper I for details). The archival data were then combined with our observations, using the {\sc casa} task {\tt concat}. A continuum-subtracted CO data cube of the combined dataset was then produced using the {\tt clean} task with \textit{Briggs} weighting (robust$=0.5$), and a final channel width of 22~km~s$^{-1}$. The sensitivity (determined in line-free channels) is 0.7~mJy, and the achieved angular resolution is 0.44$"$ (about 1.4 times higher than that of our observations alone; Table~\ref{tab:summary}). The analysis reported in the following is carried out using this combined data cube.

A summary of some relevant properties of the six $^{12}$CO(2-1) detections, the ALMA observations and the CO channel maps used in this paper is presented in Table~\ref{tab:summary}. 

\subsection{Data products}\label{sec:products}
The 3D imaging data products (moment maps and position-velocity diagrams) used in this work were created from the cleaned, continuum-subtracted CO data cubes using the masked moment technique as described by \citet[][see also \citealt{Bosma81a,Bosma81b,Kruit82,Rupen99}]{Dame11}. In this technique, a copy of the cleaned data cube is first Gaussian-smoothed spatially (with a FWHM equal to that of the synthesised beam) and then Hanning-smoothed in velocity. A three-dimensional mask is then defined by selecting all the pixels above a fixed flux-density threshold; this threshold is chosen to recover as much flux as possible while minimising the noise. We used thresholds varying from 1.2 to 2$\sigma$, depending on the significance of the CO detection (higher threshold for noisier maps). The moment maps were then produced from the un-smoothed cubes using the masked regions only \citep[e.g.][]{Davis17}. 
The integrated intensity, mean velocity and velocity dispersion maps of the objects analysed here are presented and discussed in detail in Paper I. The mean velocity and position-velocity diagram (PVD) of NGC\,3557 as extracted from the combined data cube (see Section~\ref{sec:archival_data} for details) are shown later in Figure~\ref{fig:NGC3557}.

\begin{table*}
\begin{scriptsize}
\centering
\caption{Main properties of the ALMA $^{12}$CO(2-1) datasets used in this paper.}\label{tab:summary}
\begin{tabular}{l l c c c c c c c c c c c}
\hline
\multicolumn{1}{c}{ Radio} &
\multicolumn{1}{c}{ Host } &
\multicolumn{1}{c}{ $z$ } & 
\multicolumn{1}{c}{   MRS } & 
\multicolumn{1}{c}{   $\theta$\textsubscript{maj} } & 
\multicolumn{1}{c}{   $\theta$\textsubscript{min}  } & 
\multicolumn{1}{c}{   PA} & 
\multicolumn{1}{c}{   Scale } &
\multicolumn{1}{c}{ $\nu$\textsubscript{sky}  (v\textsubscript{cen})} &
\multicolumn{1}{c}{ rms } &
\multicolumn{1}{c}{ S/N  } &
\multicolumn{1}{c}{ $\Delta$v\textsubscript{chan} } \\ 
\multicolumn{1}{c}{ source } & 
\multicolumn{1}{c}{ galaxy } &
\multicolumn{1}{c}{ }  &  
\multicolumn{1}{c}{ } &
\multicolumn{1}{c}{ }  &
\multicolumn{1}{c}{ } &
\multicolumn{1}{c}{ }  &
\multicolumn{1}{c}{  }   &
\multicolumn{1}{c}{  }   &
\multicolumn{1}{c}{ }   &
\multicolumn{1}{c}{ }   &
\multicolumn{1}{c}{ }   \\
\multicolumn{1}{c}{ } &
\multicolumn{1}{c}{ } &
\multicolumn{1}{c}{  } &
\multicolumn{1}{c}{(kpc, arcsec) } &
\multicolumn{2}{c}{  (arcsec)} &
\multicolumn{1}{c}{ (deg)}   &
\multicolumn{1}{c}{ (pc)}   &
\multicolumn{1}{c}{   (GHz) (km s$^{-1}$) } &
\multicolumn{1}{c}{ (mJy~beam$^{-1}$)}   &
\multicolumn{1}{c}{ }   &
\multicolumn{1}{c}{ (km~s$^{-1}$)}   \\
\multicolumn{1}{c}{(1)} &
\multicolumn{1}{c}{ (2)} &
\multicolumn{1}{c}{ (3)} &
\multicolumn{1}{c}{ (4)} &
\multicolumn{1}{c}{ (5) } &
\multicolumn{1}{c}{ (6) } &
\multicolumn{1}{c}{ (7)} &
\multicolumn{1}{c}{ 8}   &
\multicolumn{1}{c}{ (9)} &
\multicolumn{1}{c}{(10) }   &
\multicolumn{1}{c}{ (11)}   &
\multicolumn{1}{c}{ (12)} \\
\hline
PKS 0007$-$325&  IC\,1531 & 0.0256 &  5.6, 10.9 &  0.7  &  0.6  & 87  & 360 & 224.7774 (7702) & 0.7 &   18  &   20  \\
PKS 0131$-$31 & NGC\,612 & 0.0298 &  6.6, 11.0 &   0.3 & 0.3 &  -75 &  180 & 223.8426  (8974) &1.3  &   14  &  20  \\
PKS 0958$-$314& NGC\,3100 & 0.0088 &  1.9, 10.6    &    0.9    &    0.7    &    -87    &    160 & 228.6299 (2484) &0.6 &  45 &  10  \\
PKS 1107$-$372& NGC\,3557$^{\ast}$ & 0.0103 &  2.5, 10.7    &    0.4    &    0.4    &    -57     &    90 &  228.2319 (2999) &0.7  &   21  &  22 \\
PKS 1333$-$33 &  IC\,4296 & 0.0125 &  2.8, 10.8    &    0.6    &     0.6    &    -84    &    150 & 227.7110  (3705) & 0.2  &  8   &   40  \\
PKS 2128$-$388& NGC\,7075 & 0.0185 & 4.1, 10.8    &    0.6    &     0.6    &    -76     &    230 & 226.4196  (5483) & 0.4 &  10  &   40  \\
\hline
\end{tabular}
\parbox[t]{1\textwidth}{ \textit{Notes.} Columns: (1) Name of the radio source. (2) Host galaxy name. (3) Galaxy redshift from the NASA/IPAC extragalactic database (NED). (4) Maximum recoverable scale in kiloparsec for the array configuration, and corresponding scale in arcseconds. (5) Major axis FWHM of the synthesized beam. (6) Minor axis FWHM of the synthesized beam. (7) Position angle of the synthesized beam. (8) Spatial scale corresponding to the major axis FWHM of the synthesized beam.  (9) \co\ redshifted (sky) centre frequency estimated using the redshift listed in column (3); the corresponding velocity (v\textsubscript{cen}; LSRK system, optical convention) is reported in parentheses. (10) 1$\sigma$ rms noise level of the CO channel map measured in line-free channels at the channel width listed in column (12). (11) Peak signal-to-noise ratio of the detection. (12) Final channel width of the data cube (km\,s$^{-1}$ in the source frame).\\
$^{\ast}$The parameters listed here refer to the combined Cycle 3 and Cycle 6 data of NGC\,3557 (see Section~\ref{sec:archival_data} for details).}
\end{scriptsize}
\end{table*}

\section{Kinematic modelling: general description}\label{sec:method}
We analyse the kinematics of the six CO detections by adopting a forward-modelling approach, using the publicly available \textsc{KINematic Molecular Simulation} tool \citep[KinMS\footnote{https://github.com/TimothyADavis/KinMS};][]{Davis13}. This routine allows us to input guesses for the gas distribution and kinematics, and produces mock ALMA cubes with the same beam, pixel size and channel width as the observed ones, taking into account the observational effects of disc thickness, beam smearing, gas velocity dispersion, etc.
 The simulated data cube is generated by calculating the line-of-sight projection of the circular velocity for a defined number ($10^{5}-10^{6}$) of point-like sources that represent the gas distribution. Additional velocity contributions can be added to take into account the velocity dispersion and non-circular motions. 
 
 The \textsc{KinMS} routines are coupled with the Markov Chain Monte Carlo (MCMC) code 
 (\textsc{KinMS\_MCMC}\footnote{https://github.com/TimothyADavis/KinMS\_MCMC}), that fits the data and outputs the full Bayesian posterior probability distribution, together with the best-fitting model parameter.
 
 The purpose of carrying out the CO 3D modelling is to investigate the dynamical state of the molecular gas and derive its physical parameters. In particular, we are interested in searching for kinematical signatures that can be related to the AGN, such as deviations from circular motions (i.e.\,inflow/outflow). To this end, we initially adopt for all the sources the simple geometrical model described in the next sub-section. 
 
\subsection{Gas distribution and kinematics}\label{sec:gas_distrib_kinem}
One of the inputs of the \textsc{KinMS} models is an arbitrary parametric function that describes the surface brightness of the gas disc as a function of the radius. At the resolution of our ALMA observations, the gas surface brightness profiles turn out to be reasonably well described by a Gaussian function. The Gaussian centre and width are left free to vary in the \textsc{KinMS\_MCMC} fit. Additional free parameters of the gas disc are the integrated flux, position angle (PA), inclination ($i$), and kinematic centre (in RA, Dec and velocity). The inclination and position angles are initially fitted as single values throughout the disc. The CO disc is assumed to be thin and the model axisymmetric.

We fit the CO rotation curves using the simplest model that provides a good fit to most rotation curves on $\sim$kpc scales \citep[e.g.][]{Swinbank12,Voort18}. We assume that it follows an arctangent function of the form:
\begin{eqnarray}\label{eq:arctan}
v\textsubscript{rad}=\dfrac{2v\textsubscript{flat}}{\pi}\arctan \left(\dfrac{R}{r\textsubscript{turn}}\right)
\end{eqnarray}
where $v$\textsubscript{flat} is the asymptotic (or maximum) circular velocity in the flat part of the rotation curve, $R$ is the radius, and $r$\textsubscript{turn} is the effective radius at which the rotation curve turns over \citep[e.g.][]{Voort18}. Both $v$\textsubscript{flat} and $r$\textsubscript{turn} are left as free parameters in the fitting process, in order to find the best match to the observed gas velocity curves. 

As a first step, we always assume that the gas is in purely circular motion (i.e.\,the rotation velocity varies only radially). This approach allows us to identify the possible presence of non-circular motions, if significant residuals (i.e.\,larger than the channel width) are found between the data and the model velocity fields. Another free parameter is the internal velocity dispersion of the gas ($\sigma$\textsubscript{gas}), assumed to be spatially constant. 

\subsection{Fitting process}
To fit the model to the data we use the \textsc{KinMS\_MCMC} code, which utilises Gibbs sampling and adaptive stepping to explore the parameter space. To measure the goodness-of-fit of the model to the data we utilise a log-likelihood based on the chi-squared statistic: 
\begin{eqnarray}
\chi^{2} = \sum_{i} \left(\dfrac{{\rm data}_{i}-{\rm model}_{i}}{\sigma_{i}} \right)^{2} = \dfrac{1}{\sigma^{2}} \sum_{i} ({\rm data}_{i}-{\rm model}_{i})^{2}
\end{eqnarray}
where the sum is performed over all the pixels within the region of the data cube that the model fits, and $\sigma$ is the rms noise level measured in line-free channels of the data cube (see Table~\ref{tab:summary}), assumed to be constant for all the pixels. 
The posterior distribution of each model is then described by the log-likelihood function $\ln {\rm P} = - \chi^{2}/2$. As in \citet{Smith19}, we re-scale the uncertainty in the cube by a factor of $(2N)^{0.25}$, where $N$ is the number of pixels with detected emission in the mask, as defined in Section \ref{sec:obs}. This ensures that the uncertainty in the $\chi^2$ statistics for high N does not lead to unrealistically small uncertainties in our final fit values.

To ensure our kinematic fitting process converges, we set reasonable priors for the physical parameters that are left free to vary in the fit. The CO central velocity offset (i.e.\,the shift with respect to the kinematic centre) is allowed to vary within $\pm$five times the channel widths of each data cube (listed in Table~\ref{tab:summary}). The gas velocity dispersion ($\sigma_{\rm gas}$) is constrained to be less than the maximum line-of-sight velocity width presented in Paper I. The kinematic PA is allowed to vary by $\pm$20 degrees around the values estimated in Paper I (Table~7). The disc inclination ($\theta_{\rm inc}$) is initially left free to vary over the full physical range ($0^{\circ}-90^{\circ}$); in subsequent iterations, it is constrained to vary within $\pm20^{\circ}$ of the value at which the first chain converges. The CO maximum circular velocity (v$_{\rm flat}$) and the turnover radius (r$_{\rm turn}$) are constrained to lie within $\pm40$~km~s$^{-1}$ and two beam widths, respectively, around the values determined by visually inspecting the position-velocity diagrams (PVDs) of the six CO discs. 

Initially, the step size in each fit is adaptively scaled to ensure a minimum acceptance fraction and the chain converges. Once the \textsc{MCMC} chains converged, we re-run the entire chain for an additional $10^{5}$ steps to produce the full final posterior probability distribution. For each model parameter these probability surfaces are then marginalised over to produce a best-fit value (median of the marginalised posterior distribution) and associated 68 and 99\% confidence levels (CL).

\begin{figure*}
\centering
\begin{subfigure}[t]{0.3\textheight}
\centering
 \caption{Data moment one}\label{fig:ic1531_mom1}
\includegraphics[scale=0.34]{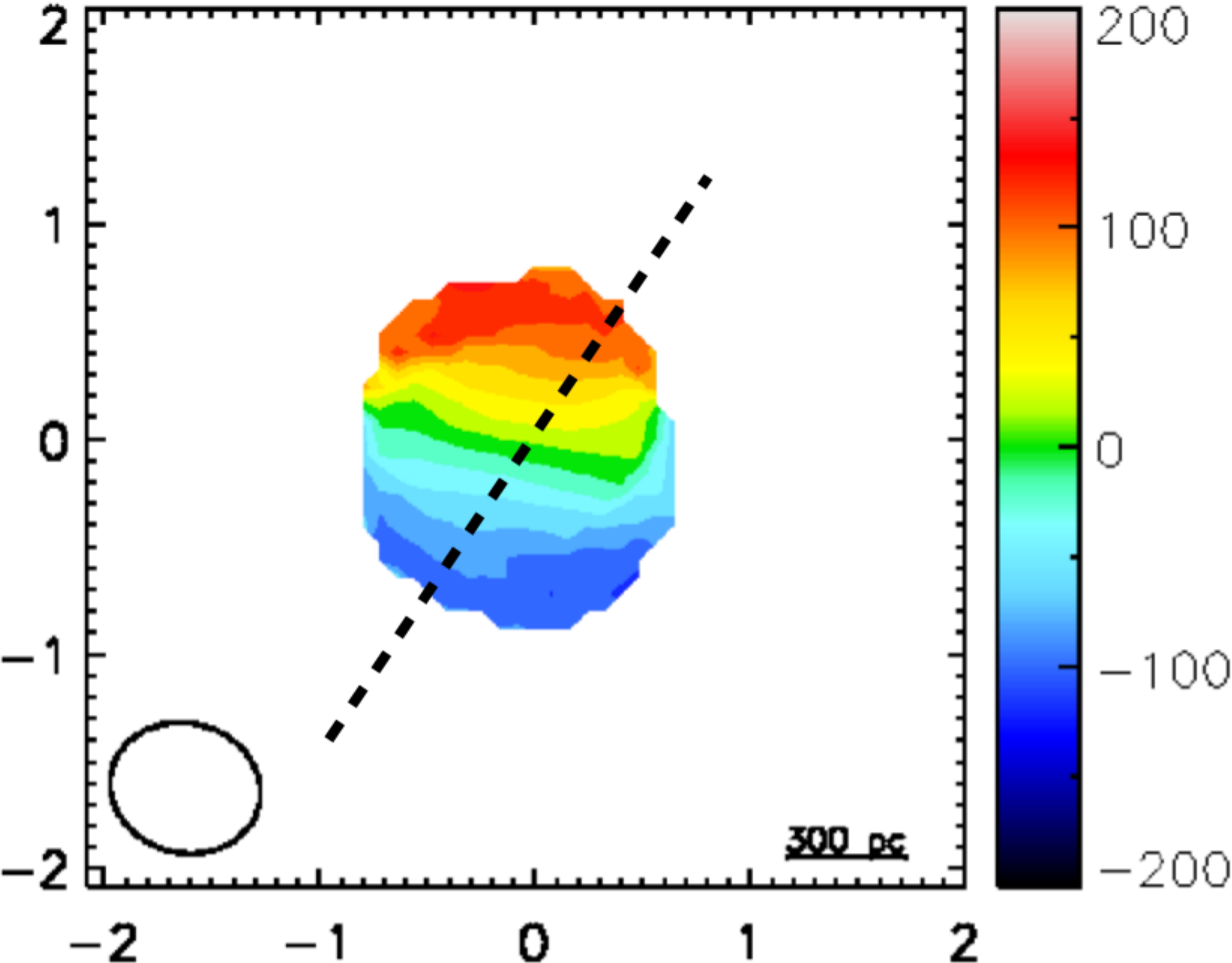}
\end{subfigure}
\hspace{10mm}
\begin{subfigure}[t]{0.3\textheight}
\centering
\caption{Model moment one}\label{fig:ic1531_mom1_mod}
\includegraphics[scale=0.3]{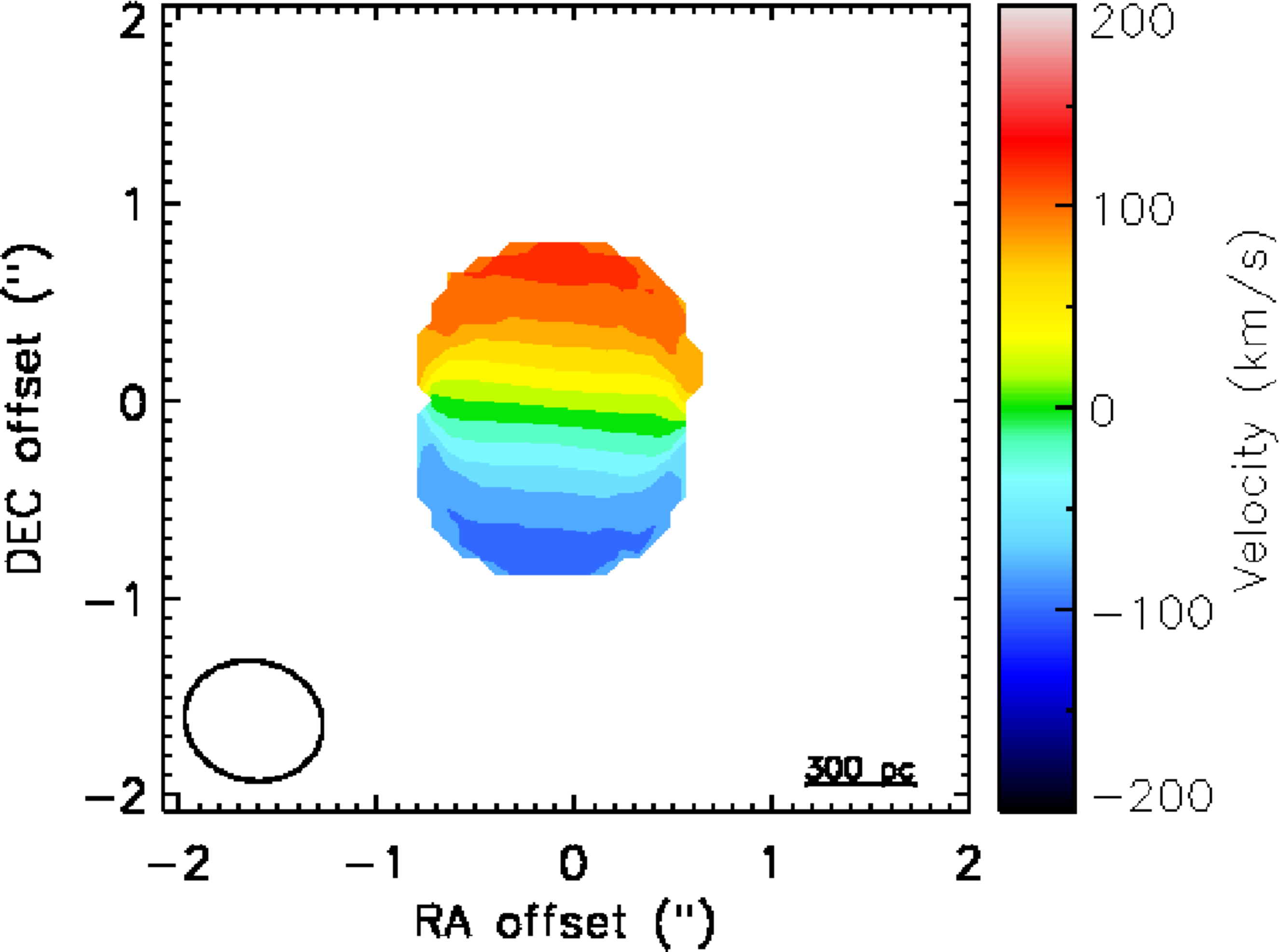}
\end{subfigure}

\medskip

\begin{subfigure}[t]{0.3\textheight}
\centering
\vspace{0pt}
\caption{Data - model moment one}\label{fig:ic1531_mom1_res}
\includegraphics[scale=0.3]{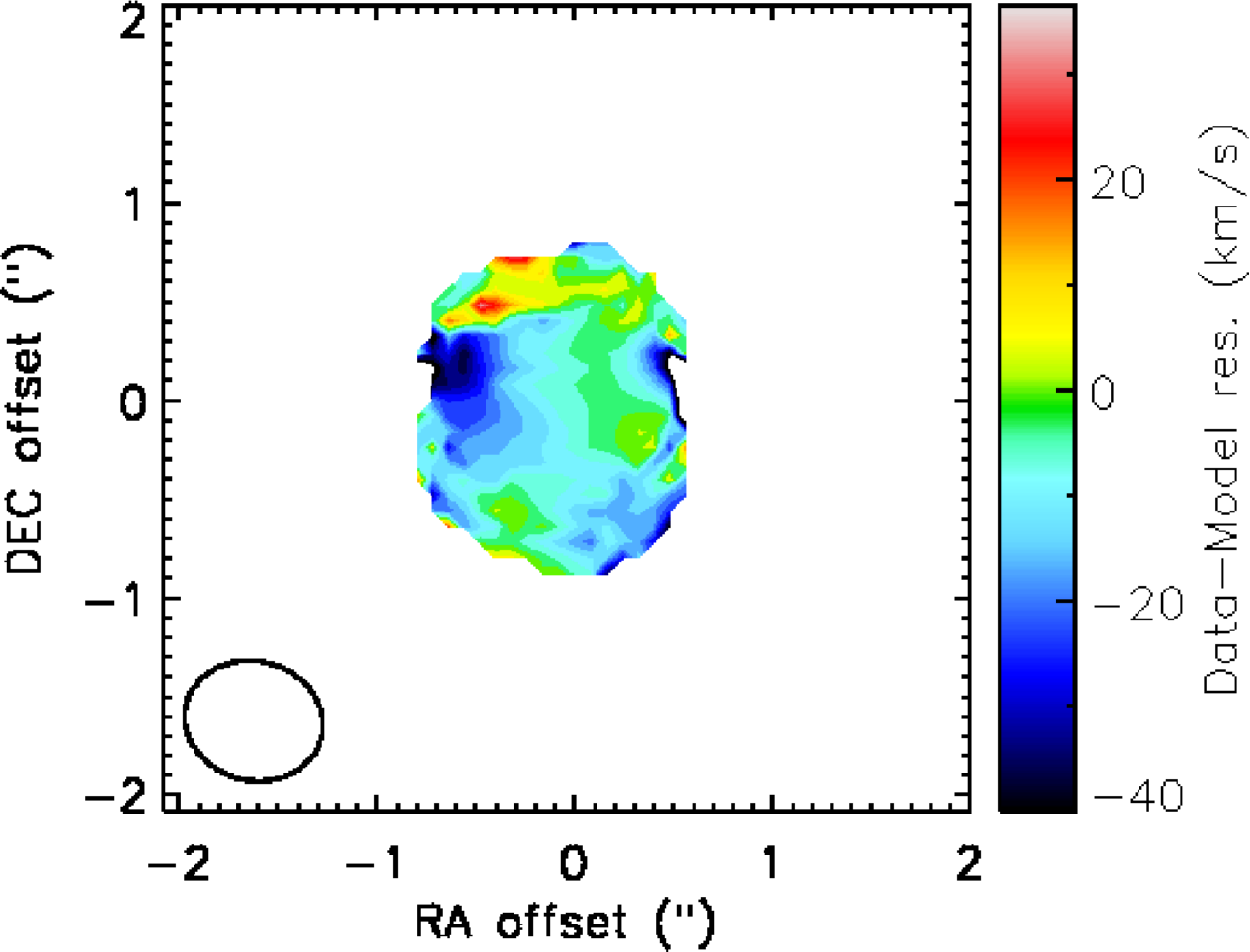}
\end{subfigure}
\hspace{10mm}
\begin{subfigure}[t]{0.3\textheight}
\centering
\vspace{0pt}
\caption{Position-velocity diagram}\label{fig:ic1531_PVD}
\includegraphics[scale=0.3]{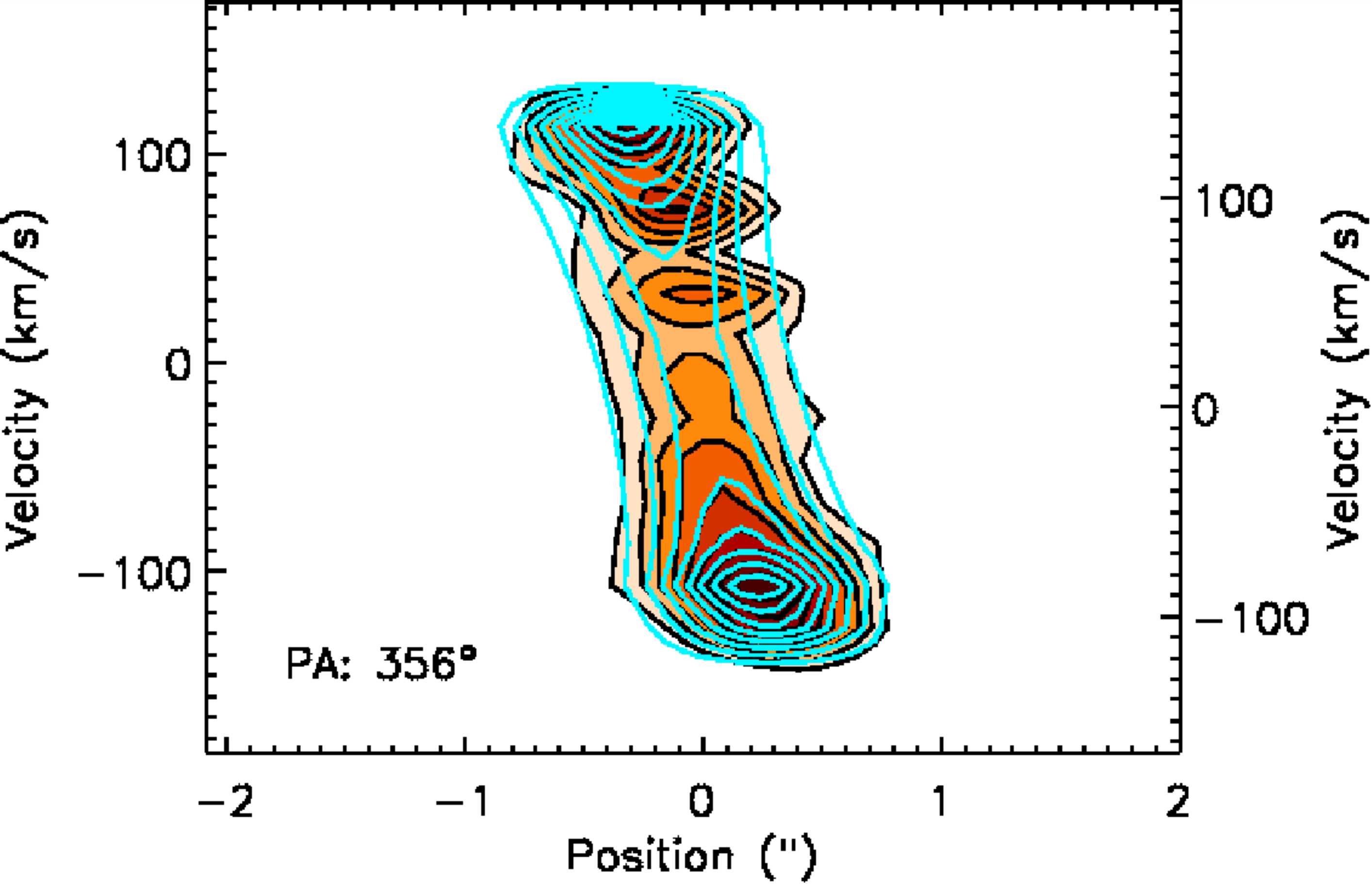}
\end{subfigure}
\caption{\textbf{IC\,1531} observed, model and residual (data-model) mean velocity maps (panels \textbf{a}, \textbf{b} and \textbf{c}, respectively). The black dashed line in panel \textbf{a} indicates the direction of the radio jet axis. The synthesised beam is shown in the bottom-left corner of each panel. The wedges to the right show the colour scale. East is to the left and North to the top. Velocities are measured in the source frame and the zero-point corresponds to the intensity-weighted centroid of the CO emission ($v_{\rm CO}$; see Table~5 of Paper I). The maps are created with the masked moment technique described in Section~\ref{sec:products}, using a data cube with a channel width of 20~km~s$^{-1}$. CO PVD (panel \textbf{d}) extracted within a rectangular area whose long axis is orientated along the kinematic position angle (indicated in the bottom-left corner of the panel) and which includes all of the Co emission along the major axis. The contours of the best-fit model are overlaid in cyan. The x-axis indicates the position offset along the extraction axis. The y-axes indicate the velocities centred on the value reported in column (7) of Table~\ref{tab:first_model} and that reported in column (9) of Table~\ref{tab:summary} (left and right right axis, respectively). The contour levels are drawn at 1, 3, 9...times the 1$\sigma$ rms noise level (see Table~\ref{tab:summary}).}\label{fig:IC1531}
\end{figure*}

\begin{figure*}
\centering
\begin{subfigure}[t]{0.3\textheight}
\centering
 \caption{Data moment one}\label{fig:ngc612_hole_mom1}
\includegraphics[scale=0.38]{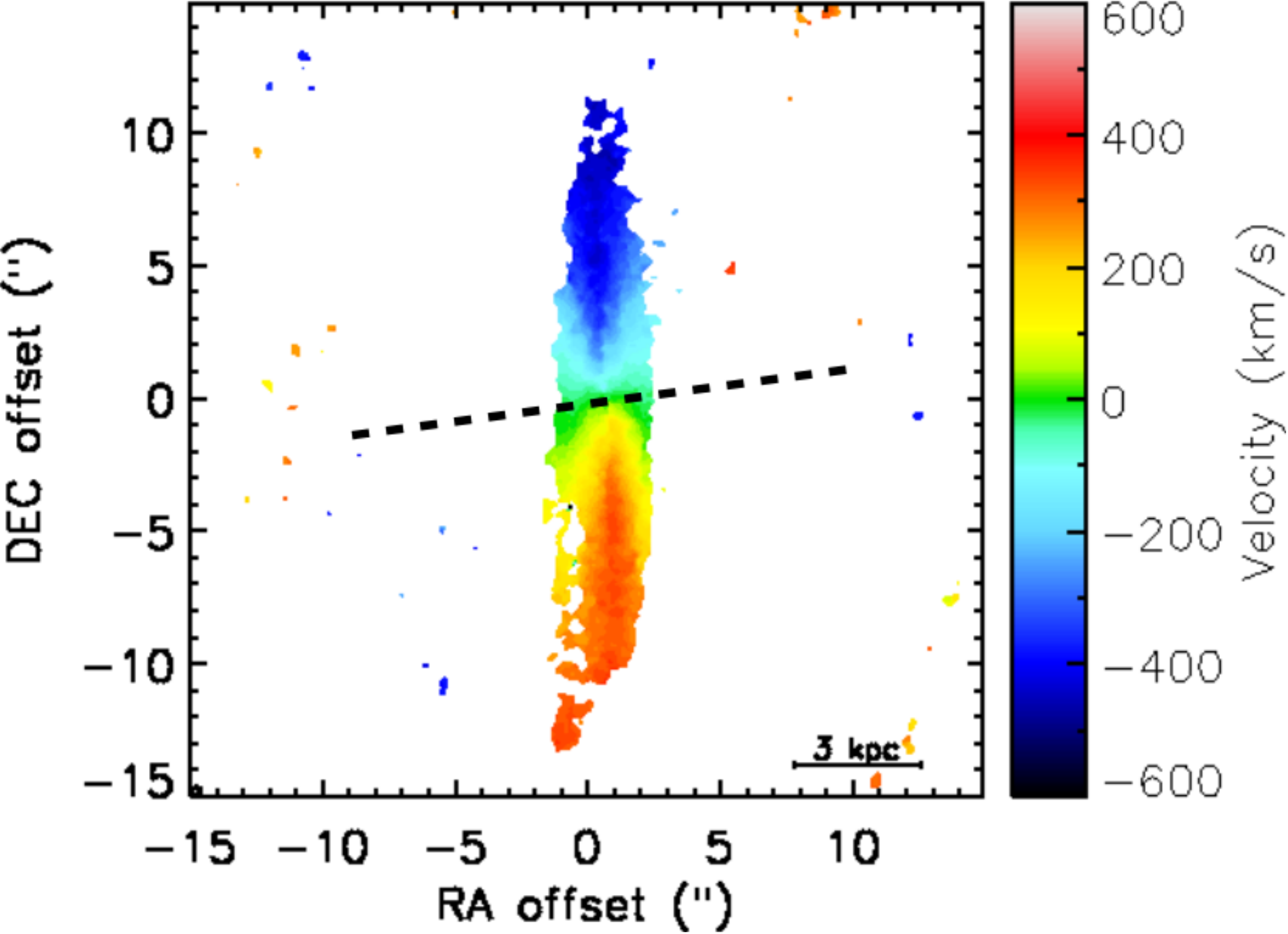}
\end{subfigure}
\hspace{10mm}
\begin{subfigure}[t]{0.3\textheight}
\centering
\caption {Model moment one}\label{fig:ngc612_hole_mom1_mod}
\includegraphics[scale=0.3]{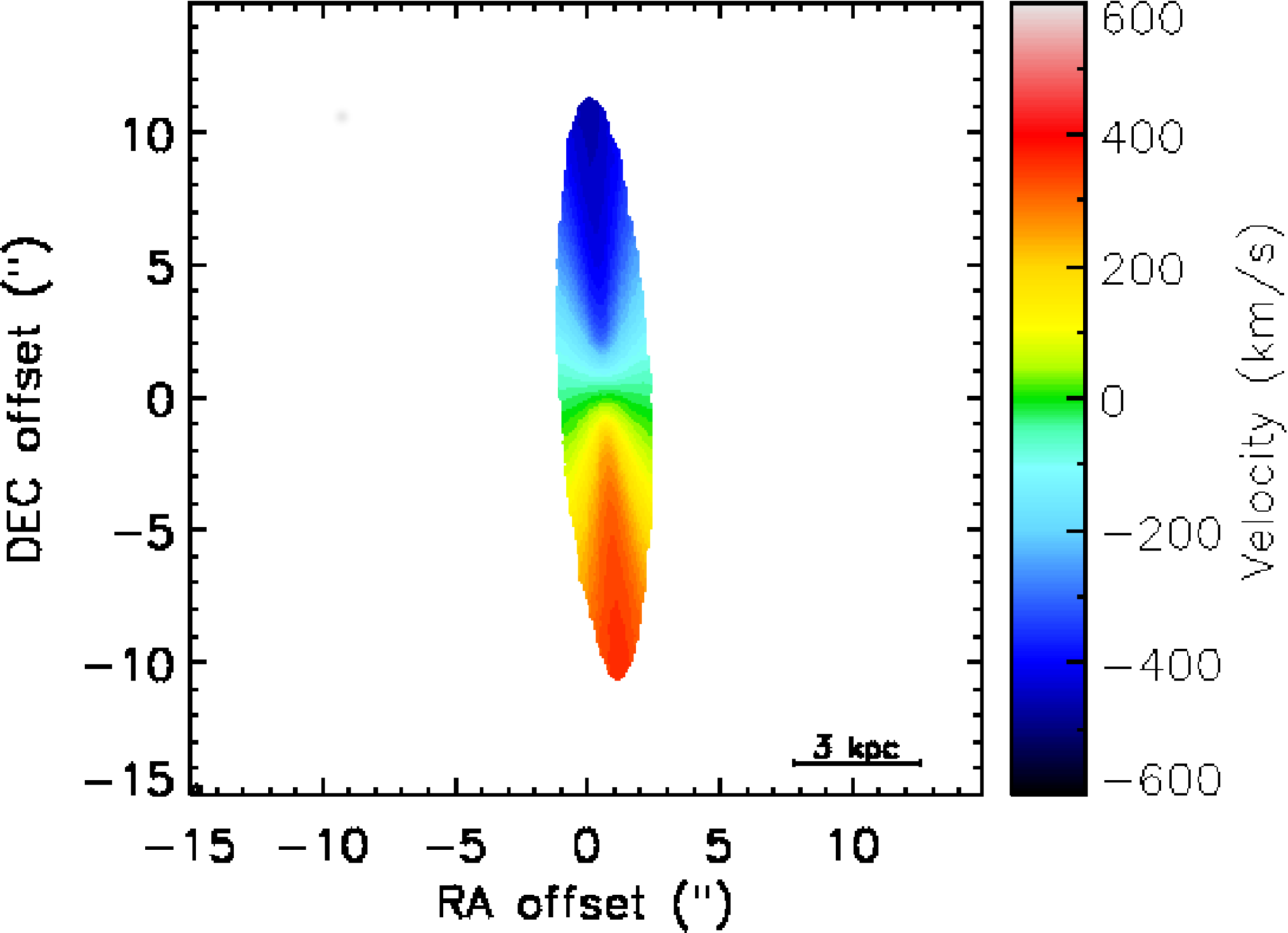}
\end{subfigure}

\medskip

\begin{subfigure}[t]{0.3\textheight}
\centering
\caption{Data - model moment one}\label{fig:ngc612_hole_mom1_res}
\includegraphics[scale=0.3]{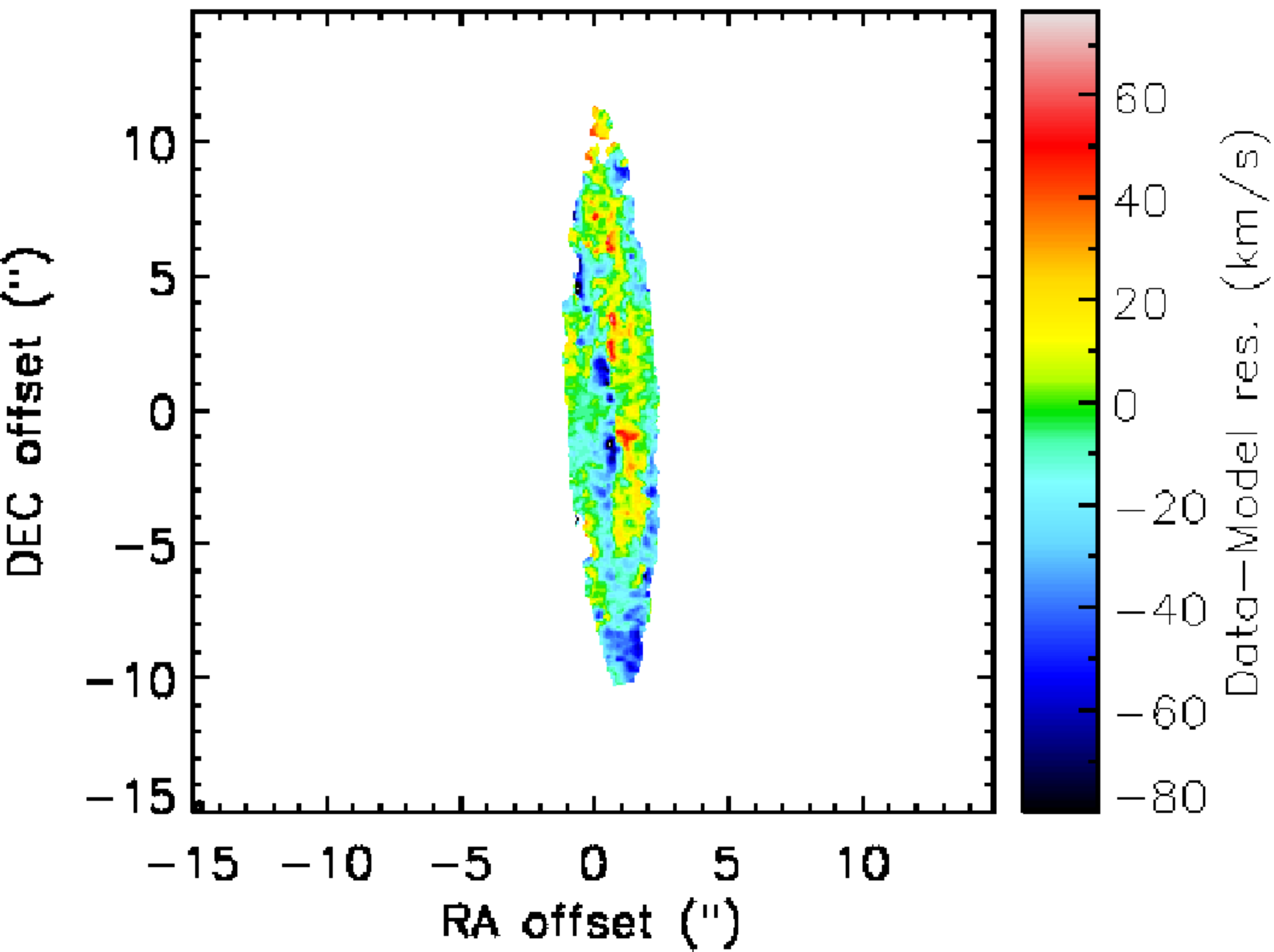}
\end{subfigure}
\hspace{10mm}
\begin{subfigure}[t]{0.3\textheight}
\centering
\caption{Position-velocity diagram}\label{fig:ngc612_hole_PVD}
\includegraphics[scale=0.3]{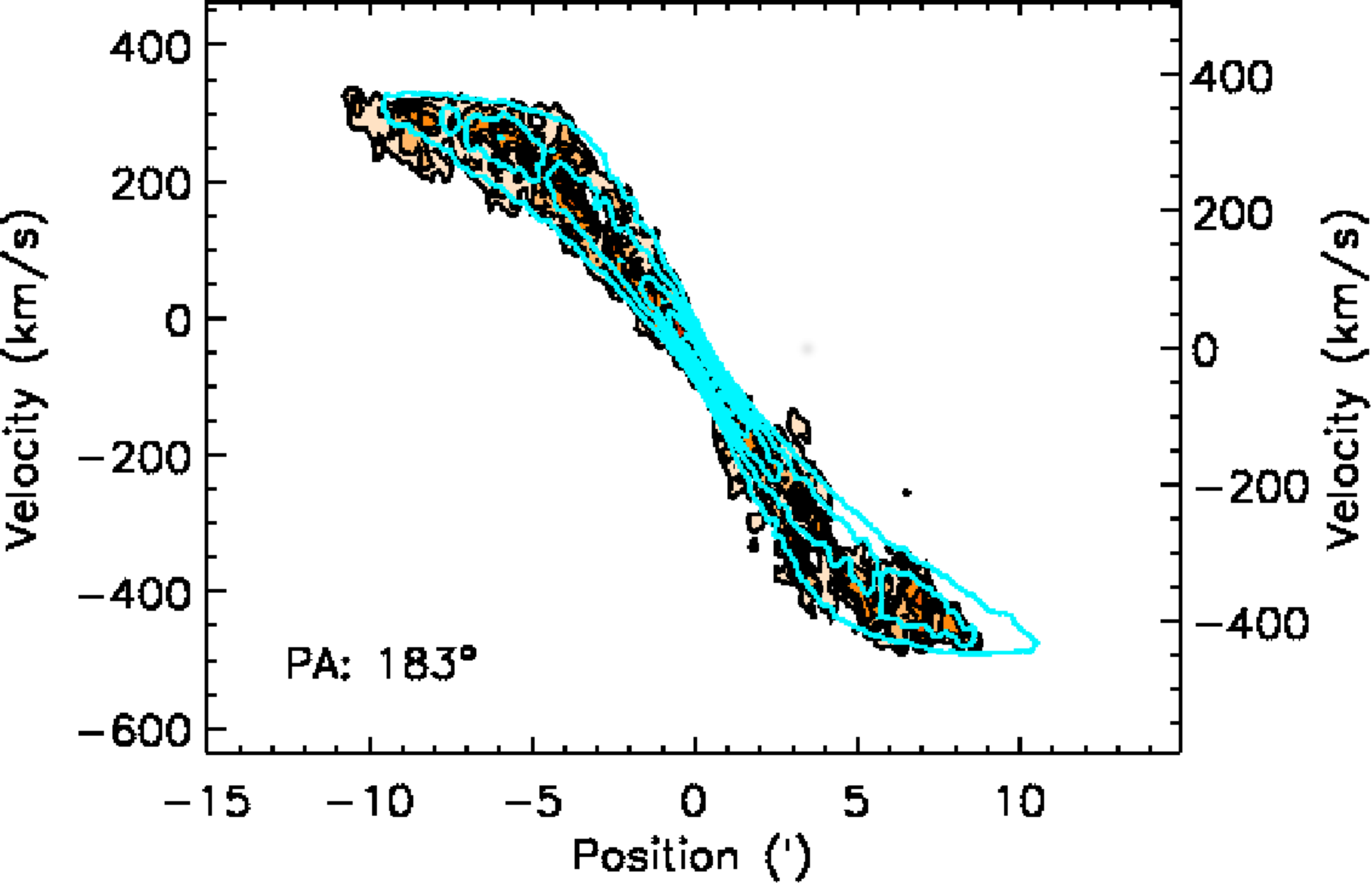}
\end{subfigure}
\caption{\textbf{NGC\,612} observed, model, residual mean velocity maps and PVD as in Figure~\ref{fig:IC1531}, created using a data cube with a channel width of 20 km~s$^{-1}$.}\label{fig:NGC612}
\end{figure*}

\begin{figure*}
\centering
\begin{subfigure}[t]{0.3\textheight}
\centering
 \caption{Data moment one}\label{fig:IC4296_mom1}
\includegraphics[scale=0.37]{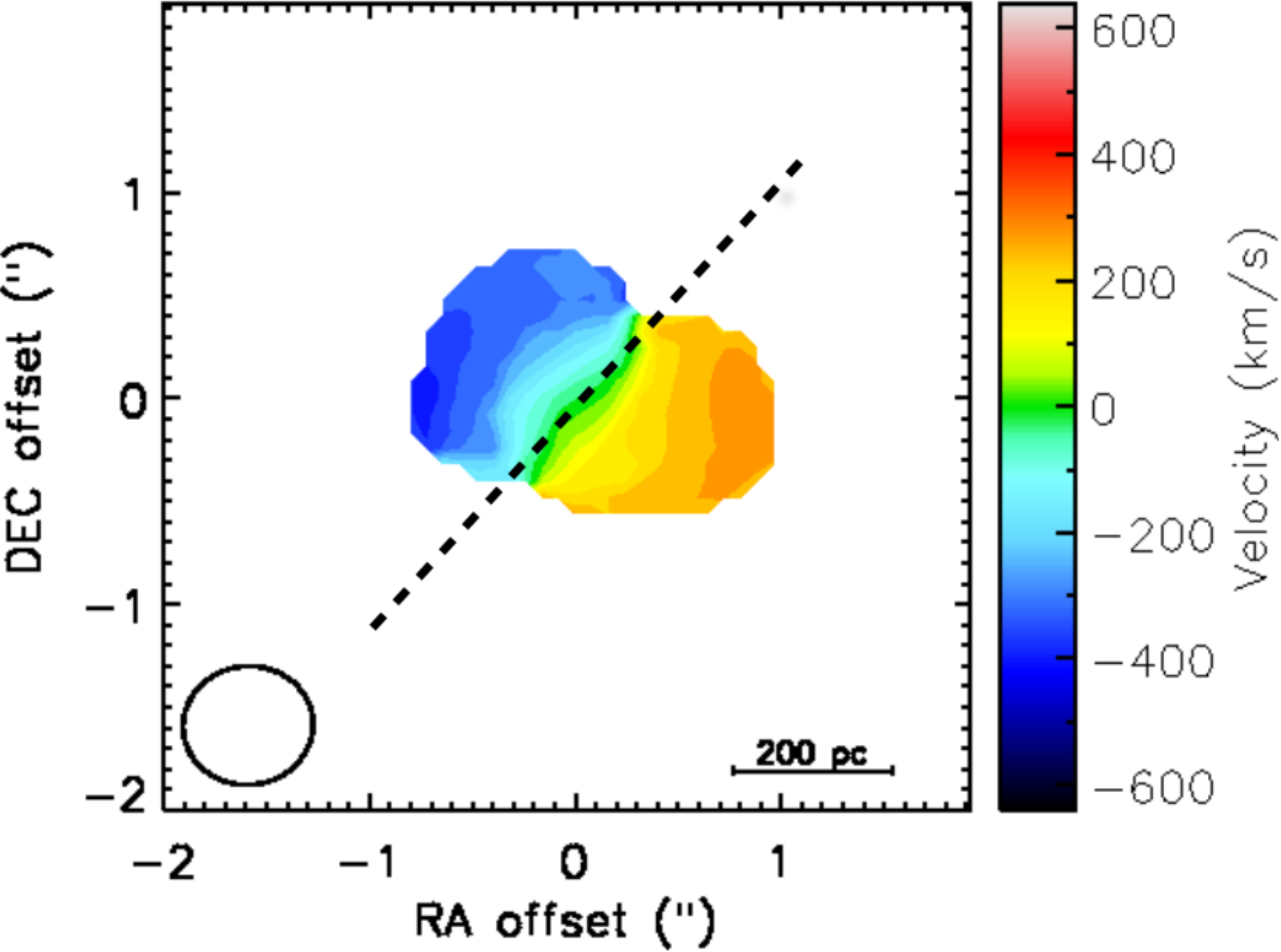}
\end{subfigure}
\hspace{10mm}
\begin{subfigure}[t]{0.3\textheight}
\centering
\caption{Model moment one}\label{fig:IC4296_mom1_mod}
\includegraphics[scale=0.3]{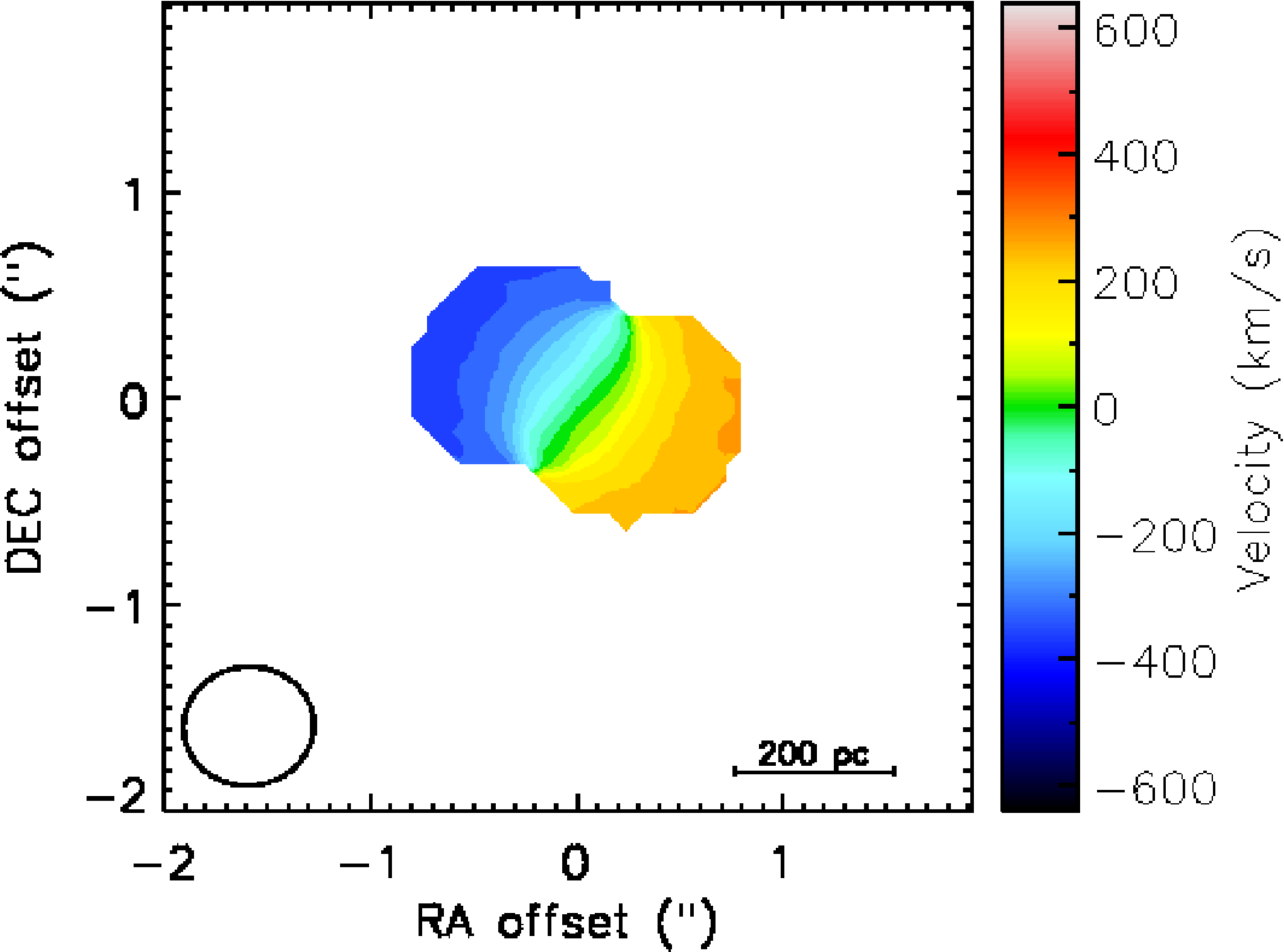}
\end{subfigure}

\medskip

\begin{subfigure}[t]{0.3\textheight}
\centering
\vspace{0pt}
\caption{Data - model moment one}\label{fig:IC4296_mom1_res}
\includegraphics[scale=0.3]{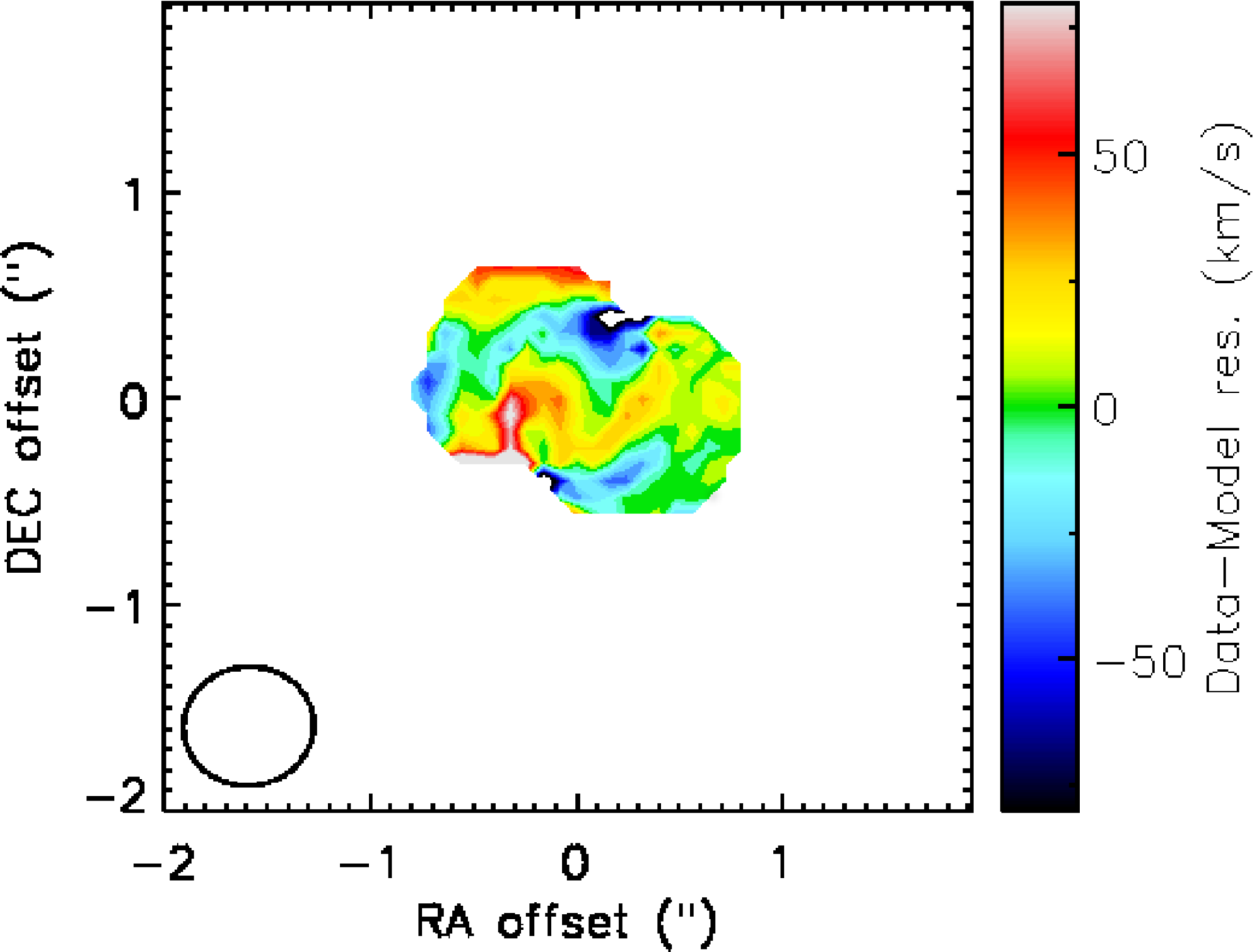}
\end{subfigure}
\hspace{10mm}
\begin{subfigure}[t]{0.3\textheight}
\centering
\vspace{0pt}
\caption{Position-velocity diagram}\label{fig:IC4296_PVD}
\includegraphics[scale=0.3]{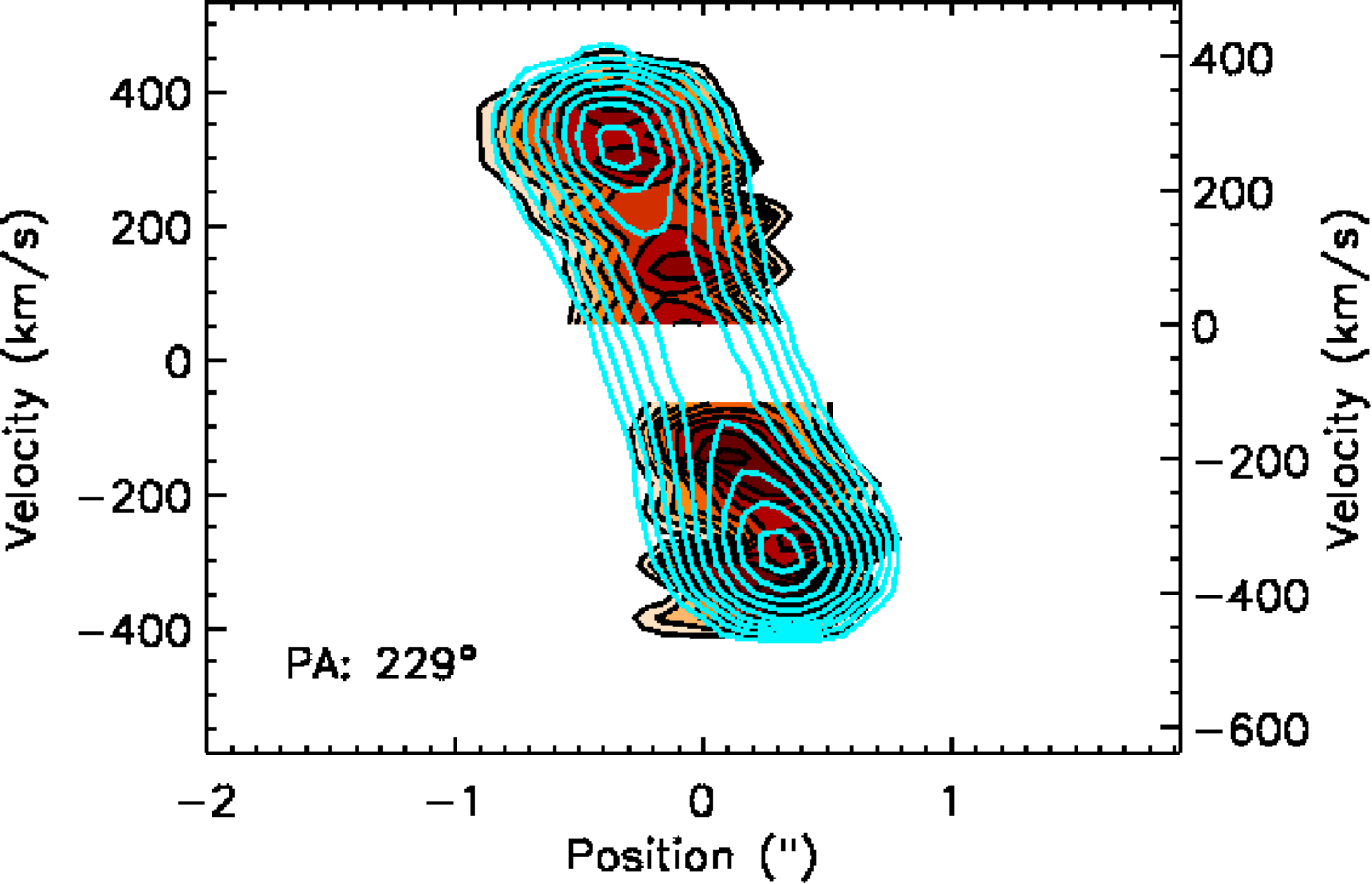}
\end{subfigure}
\caption{\textbf{IC\,4296} observed, model, residual mean velocity maps and PVD as in Figure~\ref{fig:IC1531}, created using a data cube with a channel width of 40 km~s$^{-1}$.}\label{fig:IC4296}
\end{figure*}

\begin{figure*}
\centering
\begin{subfigure}[t]{0.3\textheight}
\centering
 \caption{Data moment one}\label{fig:NGC7075_mom1}
\includegraphics[scale=0.4]{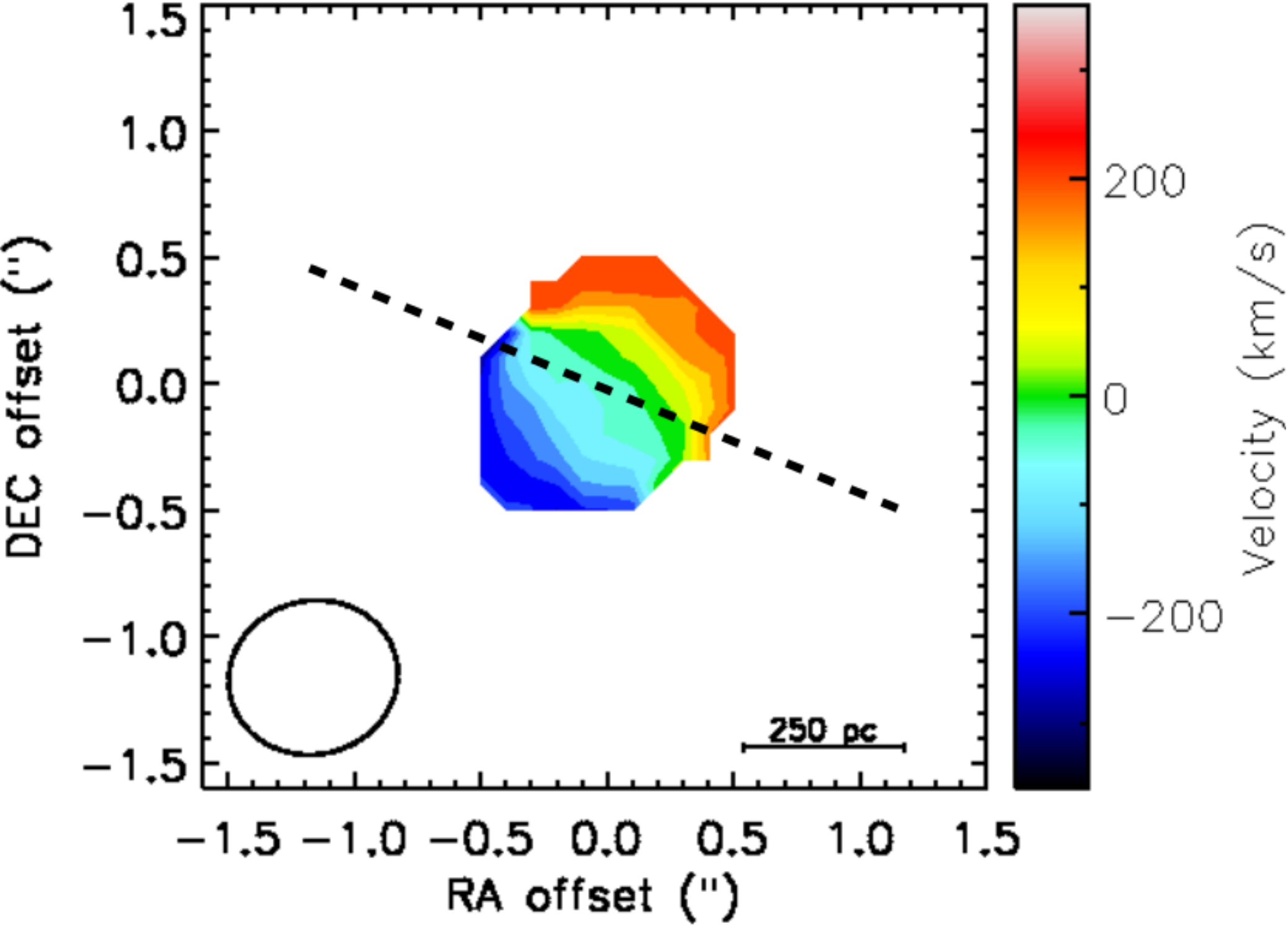}
\end{subfigure}
\hspace{15mm}
\begin{subfigure}[t]{0.3\textheight}
\centering
\caption{Model moment one}\label{fig:NGC7075_mom1_mod}
\includegraphics[scale=0.3]{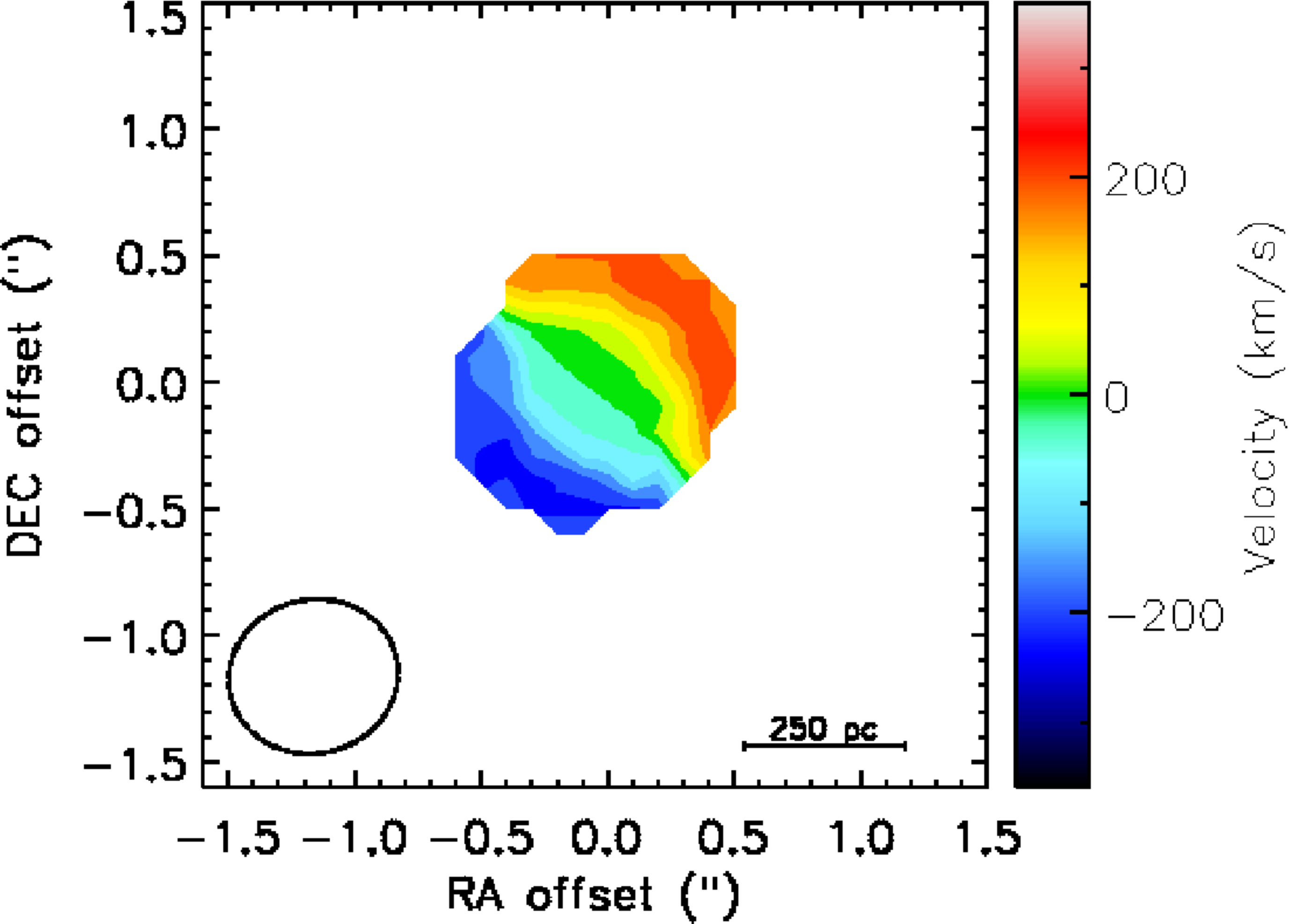}
\end{subfigure}

\medskip

\begin{subfigure}[t]{0.3\textheight}
\centering
\vspace{0pt}
\caption{Data - model moment one}\label{fig:NGC7075_mom1_res}
\includegraphics[scale=0.3]{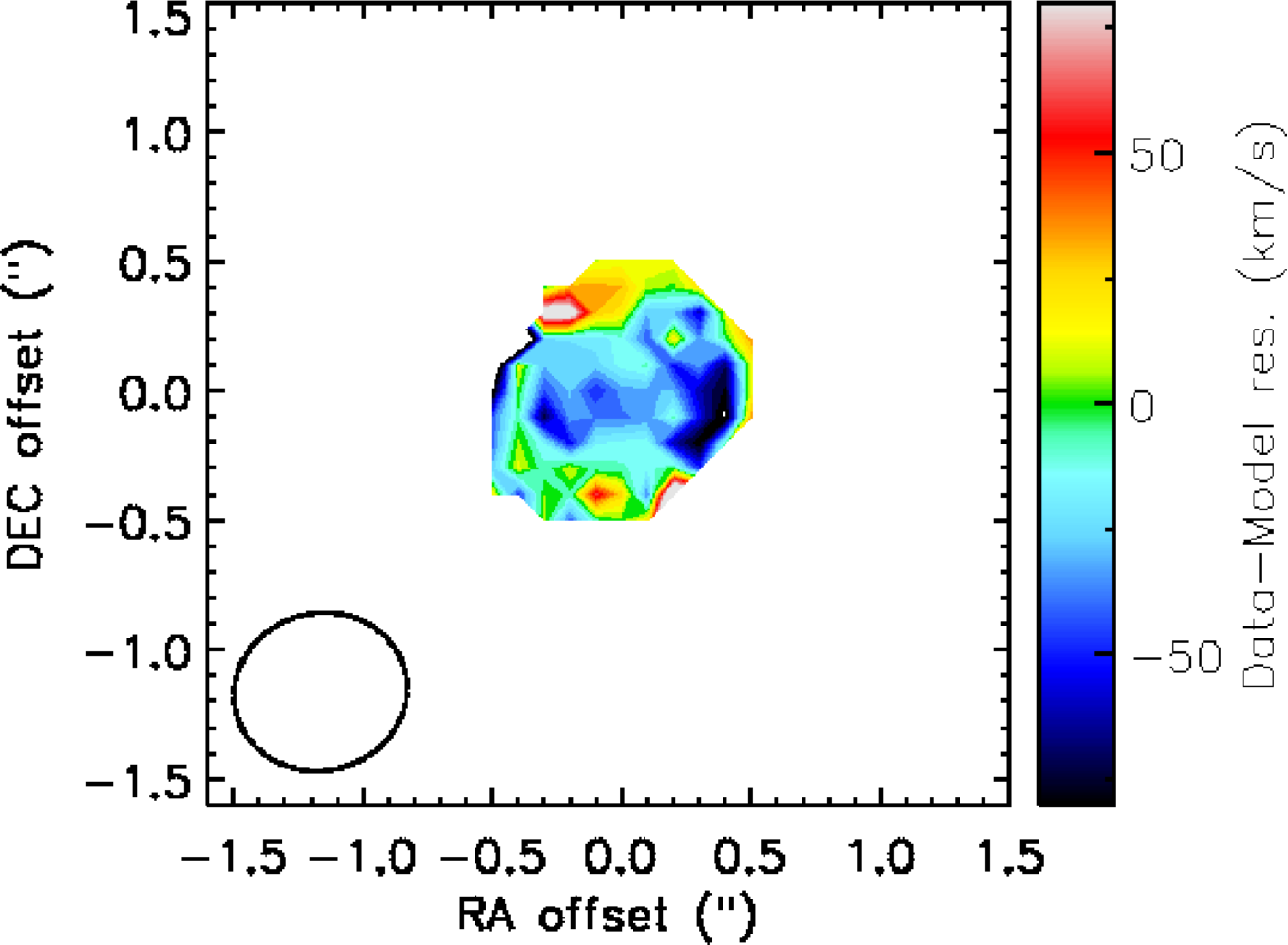}
\end{subfigure}
\hspace{10mm}
\begin{subfigure}[t]{0.3\textheight}
\centering
\vspace{0pt}
\caption{Position-velocity diagram}\label{fig:NGC7075_PVD}
\includegraphics[scale=0.3]{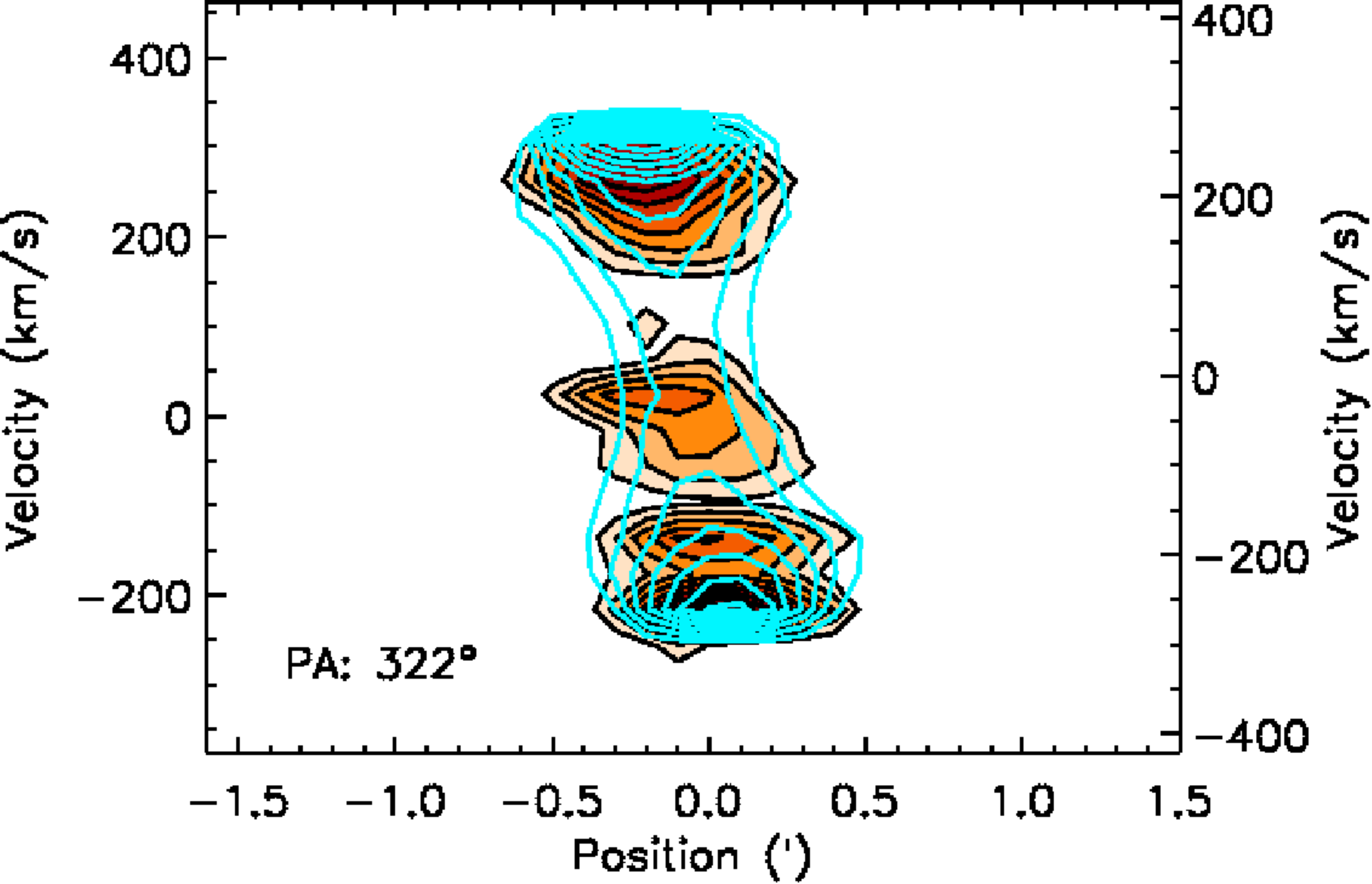}
\end{subfigure}
\caption{\textbf{NGC\,7075} observed, model, residual mean velocity maps and PVD as in Figure~\ref{fig:IC1531}, created using a data cube with a channel width of 40 km~s$^{-1}$.}\label{fig:NGC7075}
\end{figure*}

\section{Kinematic modelling: individual sources}\label{sec:individual_sources}
 At the resolution of our ALMA observations (Table~\ref{tab:summary}), the majority of the CO discs are only marginally resolved, so that the simple geometrical model described above reproduces reasonably well the bulk of the observed gas structures. Nevertheless, modifications to that basic model such as inclusion of warps or an inner surface-brightness cut-off are required in some cases to reproduce the gas distribution and kinematics accurately. In Sections~\ref{sec_ic1531} -- \ref{sec_ngc7075}, we  describe the modelling details for the four individual sources where no or minor modifications to the model are required to get a good fit to the data. The best-fit parameters, associated errors (including only random uncertainties at 99\% confidence level), and the reduced chi-squared ($\chi^{2}_{\rm red}$)\footnote{$\chi^{2}_{\rm red}= \dfrac{\chi^{2}}{\rm d.o.f}$, where $\rm d.o.f.$ is the number of degrees of freedom for the adopted model.} of the models adopted for these four objects are listed in Table~\ref{tab:first_model}. The upper panels of Figures~\ref{fig:IC1531} $-$ \ref{fig:NGC7075} compare the observed and modelled mean velocity maps (moment 1; panels a and b). The lower panels of Figures~\ref{fig:IC1531} $-$ \ref{fig:NGC7075} show the data-model residual moment 1 map (panel c) and the major axis PVD with the best-fit model over-plotted (cyan contours; panel d). 
 
 NGC\,3557 and NGC\,3100 are special cases: the asymmetries observed both in their morphology and kinematics require a more complex approach, discussed separately in Sections~\ref{sec:NGC3557} and \ref{sec:NGC3100}.

\subsection{IC\,1531} 
\label{sec_ic1531}
The barely resolved CO disc of IC\,1531 mainly exhibits regular rotation pattern and morphology. Therefore, a simple axisymmetric model reproduces the observed gas features reasonably well (Fig.~\ref{fig:IC1531}). However, an s-shaped distortion is visible in the rotation pattern at $\approx0.5''$, along the zero-velocity iso-contour (Fig.~\ref{fig:ic1531_mom1}). Asymmetric structures are also present at the same position in the velocity curve, between $\approx0$ and $\approx100$~km~s$^{-1}$ (Fig.~\ref{fig:ic1531_PVD}). Such distortions may trace the presence of a warp and/or non-circular motions (e.g.\,inflow, outflow, streaming). We attempt to include both position and inclination angle warps in our model, but these are not favoured by the fitting routine. This may suggest that non-circular motions dominate, or that given the poor resolution of our data there was simply not enough information to constrain more complex models. Distortions not reproduced by the model probably give rise to the low-level residuals ($\approx-30$~km~s$^{-1}$) visible in Figure~\ref{fig:ic1531_mom1_res} at one edge of the gas disc, at the position of the zero-velocity contour. Nevertheless, the best-fitting model in Figure~\ref{fig:IC1531} shows no significant departures from the data, suggesting that any radial motions or asymmetries are small in amplitude and do not critically affect our analysis. Higher resolution observations are needed to investigate this object further.

\subsection{NGC\,612}
NGC\,612 is exceptional among our sources: we detect a large CO(2-1) disc extending $\approx$10~kpc along the major axis, with an estimated molecular gas mass of $2.0\times10^{10}$~M$_{\odot}$ (about two order of magnitude larger than in the other CO detections; see Paper I).

In this case, our initial model leaves large residuals in the inner part of the disc (peak amplitudes$\approx\pm80$~km~s$^{-1}$). Furthermore, the sharp edges of the velocity curve (Fig.~\ref{fig:ngc612_hole_PVD}) cannot be well reproduced.
The best-fitting model shown in Figure~\ref{fig:NGC612} is obtained assuming a ring-like CO distribution with a Gaussian surface brightness profile. The size of the central hole is left free to vary and is found to be $\approx0.3$\,arcsec (180~pc; Table~\ref{tab:first_model}). Some high-level residuals (peak amplitudes$\approx\pm60$~km~s$^{-1}$) are still visible in the central part of the gas distribution, possibly suggesting the presence of non-circular motions. Significant residuals (peak amplitudes$\approx\pm40$~km~s$^{-1}$) are also visible at the outer edges of the disc (Fig.~\ref{fig:ngc612_hole_mom1_res}) at the locations of the observed asymmetries (Fig.~\ref{fig:ngc612_hole_mom1}). These are likely to be associated with the presence of a warp. If present, this affects the disc only on the largest scales and for this reason we do not attempt to model it. 

\subsection{IC\,4296}
Although the basic model described in Section~\ref{sec:gas_distrib_kinem} provides a reasonable representation of the observations, the CO rotation pattern of IC\,4296 is well reproduced only when accounting for the presence of a warp (Fig.~\ref{fig:IC4296_mom1} and \ref{fig:IC4296_mom1_mod}). Specifically, we let the kinematic position angle vary linearly in the radial direction along the disc, finding a best-fitting range of 240 -- 220 degrees (Table~\ref{tab:first_model}). Nevertheless, low level residuals (slightly larger than the channel width$=40$~km~s$^{-1}$) are still visible along the disc minor axis (Fig.~\ref{fig:IC4296_mom1_res}), possibly suggesting the presence of non-circular motions. It is hard, however, to establish whether such motions are truly present or not, given the resolution of the current dataset (Table~\ref{tab:summary}). The gas distribution is well reproduced by this axisymmetric model, even though the gas disc is lopsided at its edges (Fig.~\ref{fig:IC4296_mom1}). We note that an absorption feature against the radio core is detected in this source (see Paper I), giving rise to the hole in the gas distribution visible in the velocity curve (Fig.~\ref{fig:IC4296_PVD}). The channels in which the CO absorption dominates are excluded from the model. 

\subsection{NGC\,7075}
\label{sec_ngc7075}
The CO disc in NGC\,7075 is barely resolved by our ALMA observations (see Paper I). Our best-fitting basic model shows no significant departures from the observations, although some residuals are present (Fig.~\ref{fig:NGC7075_mom1_res}). Interestingly, there is an asymmetry in the PVD that is not reproduced by the model (Fig.~\ref{fig:NGC7075_PVD}). Furthermore, a hole is visible in the gas distribution at positive velocities. These features indicate that some disturbance must be occurring in the gas disc (both morphologically and kinematically), but higher resolution observations would be needed to investigate the origin of the observed structures.

\begin{table*}
\centering
\caption{Best-fitting parameters for the four sources which are adequately fitted by simple models.}
\label{tab:first_model}
\begin{tabular}{l c c c c c c c c c c}
\hline
\multicolumn{1}{c}{ Target } &
\multicolumn{1}{c}{ PA} & 
\multicolumn{1}{c}{ $\theta_{\rm inc}$ } &
\multicolumn{1}{c}{ v$_{\rm flat}$ } &
\multicolumn{1}{c}{ r$_{\rm turn}$ } &
\multicolumn{1}{c}{  $\sigma_{\rm gas}$ } & 
\multicolumn{1}{c}{   v$_{\rm offset}$ (v$_{\rm cen}$)} &
\multicolumn{1}{c}{   R$_{\rm hole}$} &
\multicolumn{1}{c}{  $\chi^{2}_{\rm red}$  } \\      
\multicolumn{1}{c}{ } &       
\multicolumn{1}{c}{  (deg) } &
\multicolumn{1}{c}{  (deg) } &   
\multicolumn{1}{c}{   (km s$^{-1}$)} & 
\multicolumn{1}{c}{   (arcsec) } & 
\multicolumn{1}{c}{   (km s$^{-1}$) } &          
\multicolumn{1}{c}{   (km s$^{-1}$) (km s$^{-1}$)} & 
\multicolumn{1}{c}{ (arcsec)   } &
\multicolumn{1}{c}{    } \\
\multicolumn{1}{c}{   (1) } &   
\multicolumn{1}{c}{   (2) } &
\multicolumn{1}{c}{   (3) } &
\multicolumn{1}{c}{   (4) } & 
\multicolumn{1}{c}{   (5) } &
\multicolumn{1}{c}{   (6) } &
\multicolumn{1}{c}{   (7) } &
\multicolumn{1}{c}{   (8) } &
\multicolumn{1}{c}{   (9) } \\

\hline
 IC 1531 & 356$\pm1.0$  & 32$\pm2.5$  & 272$\pm29$ & 0.07$\pm0.02$ & 3.4$\pm2.3$ & $-$25$\pm$1.0 (7677) & $-$ & 1.2   \\ 
  NGC 612  &  183$\pm0.05$  &  81$\pm0.01$  & 453$\pm0.04$ & 1.2$\pm$0.05 & 20$\pm0.02$   &  $-$95$\pm$0.01 (8879) & 0.3 & 1.2 \\ 
 IC 4296  & 240-220 & 68$\pm1.5$   &  404$\pm$1.5 & 0.01$\pm$0.002 &  64$\pm1.5$    & $-16\pm2.0$ (3691)   & $-$ & 1.0  \\  
 NGC 7075 &   322$\pm4.0$   &  46$\pm$1.5 & 446$\pm56$ & 0.03$\pm$0.02 & 5.1$\pm2.6$  &  7.8$\pm3.0$ (5491)  & $-$ & 1.1    \\
\hline
\end{tabular}
\parbox[t]{1\textwidth}{ \textit{Notes.} $-$ Columns: (1) Target name. (2) Kinematic position angle of the CO disc (i.e.\,the PA measured counterclockwise from North to the approaching side of the velocity field). This is given as a range for IC\,4296, where a position angle war is included. (3) Inclination of the disc. (4) Asymptotic (or maximum) circular velocity in the flat part of the rotation curve. (5) Effective radius at which the rotation curve turns over. (6) Gas velocity dispersion. (7) Velocity offset of the kinematic centre (i.e.\,offset with respect to the expected and observed velocity zero-point; the expected zero-point corresponds to the redshifted frequency of the CO(2-1) transition listed in column (9) of Table~\ref{tab:summary}). The corresponding CO central velocity is reported in parentheses. (8) Radius of the central surface brightness cut-off. (9) Reduced $\chi^{2}$ of the best-fit model.}
\end{table*}

\subsection{NGC 3557 }\label{sec:NGC3557}
\begin{figure*}
\centering
\begin{subfigure}[t]{0.3\textheight}
\centering
 \caption{Data moment one}\label{fig:ngc3557_mom1}
\includegraphics[scale=0.37]{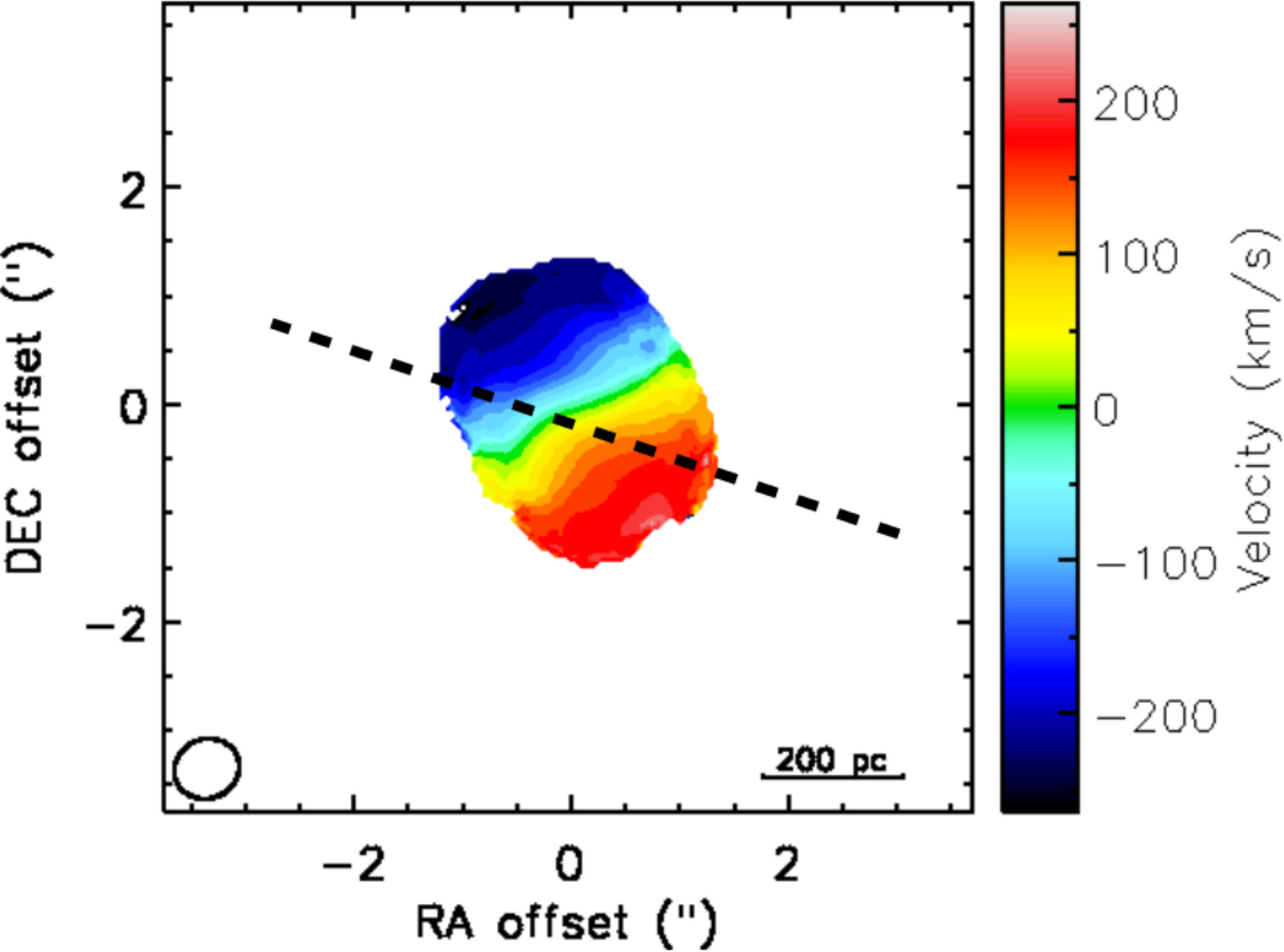}
\end{subfigure}
\hspace{10mm}
\begin{subfigure}[t]{0.3\textheight}
\centering
\caption{Model moment one}\label{fig:ngc3557_mom1_mod}
\includegraphics[scale=0.3]{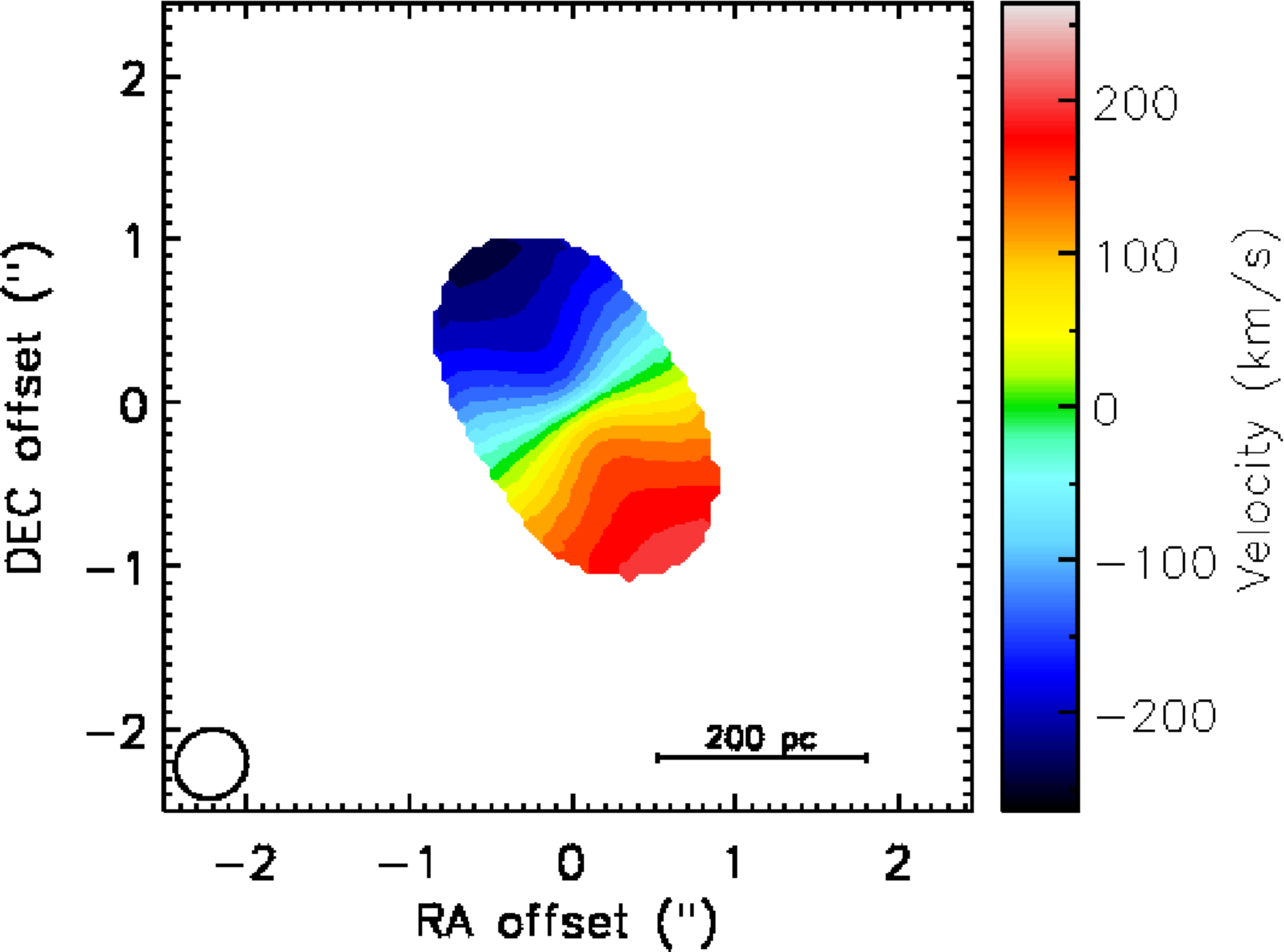}
\end{subfigure}

\medskip

\begin{subfigure}[t]{0.3\textheight}
\centering
\vspace{0pt}
\caption{Data - model moment one}\label{fig:ngc3557_mom1_res}
\includegraphics[scale=0.3]{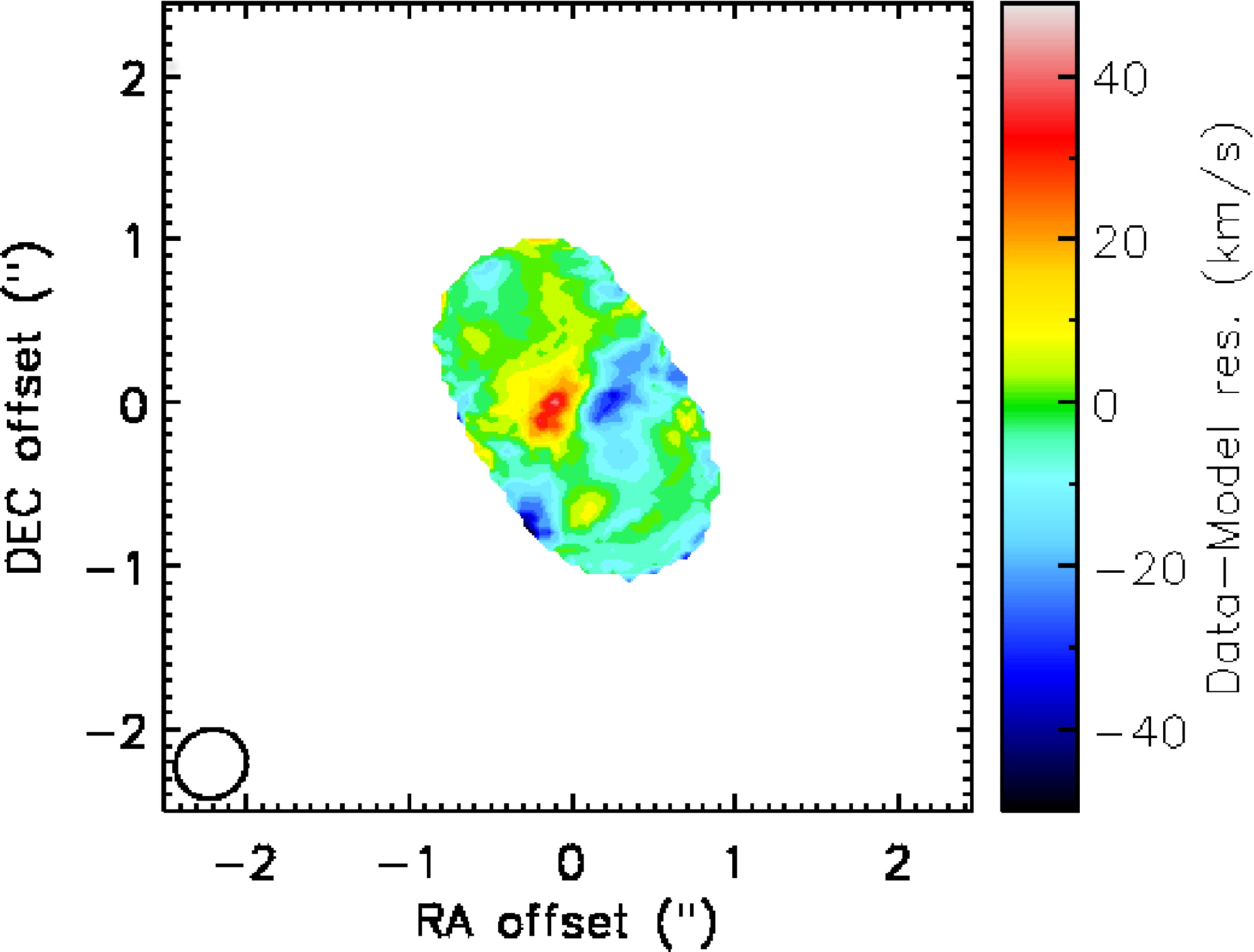}
\end{subfigure}
\hspace{10mm}
\begin{subfigure}[t]{0.3\textheight}
\centering
\vspace{0pt}
\caption{Position-velocity diagram}\label{fig:ngc3557_PVD}
\includegraphics[scale=0.3]{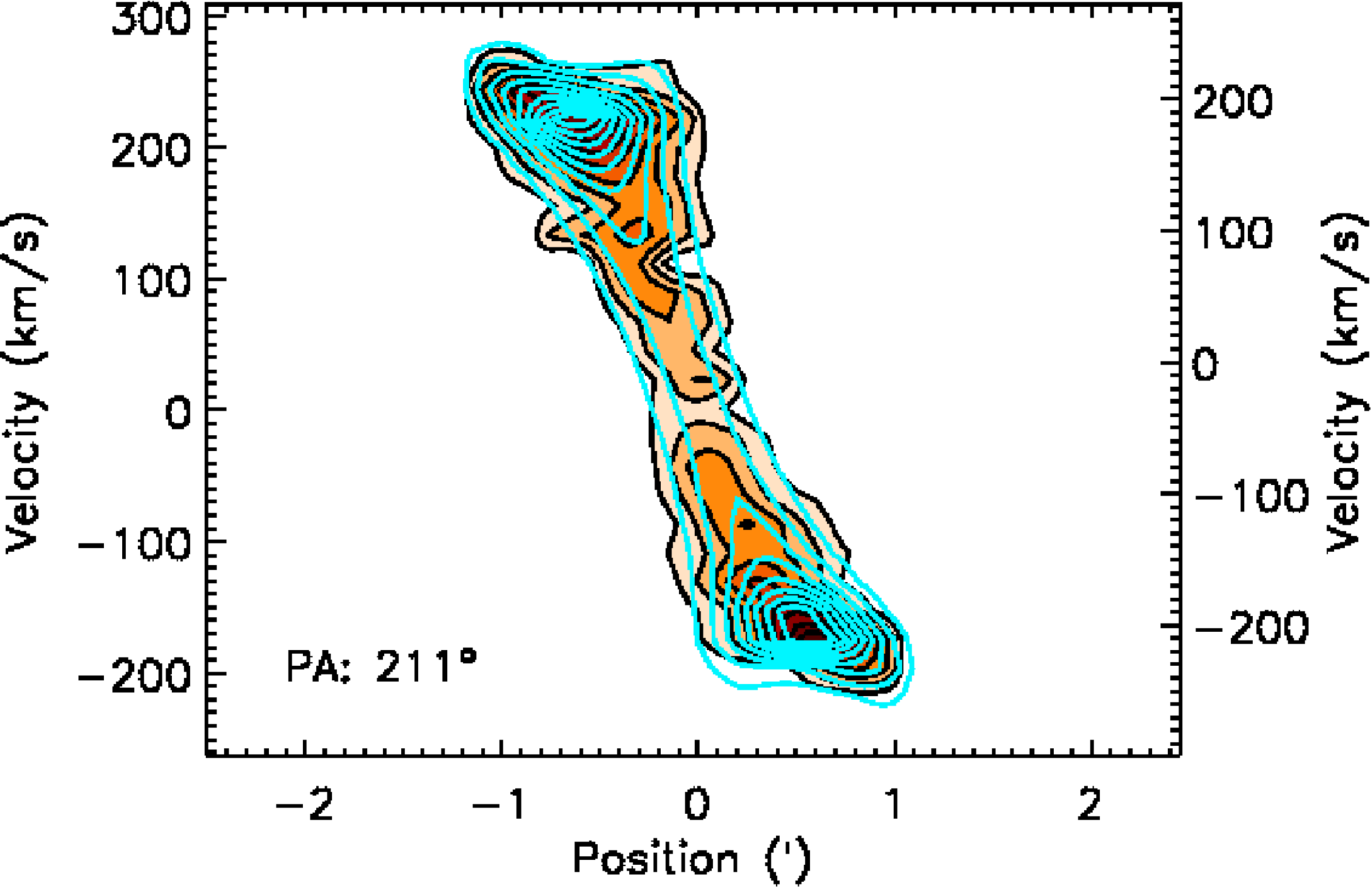}
\end{subfigure}
\caption{NGC\,3557 observed, mock, residual mean velocity maps and PVD as in Figure~\ref{fig:IC1531}, created using a data cube with a channel width of 22 km~s$^{-1}$.}\label{fig:NGC3557}
\end{figure*}

The CO(2-1) emission of NGC\,3557 shows a disc-like structure, but asymmetric features are evident both in the rotation pattern (Fig.~\ref{fig:ngc3557_mom1}) and the velocity curve (Fig.~\ref{fig:ngc3557_PVD}). Distortions are clearly visible in the mean velocity map at an RA offset of $\approx-0.8''$ (positive velocities; Fig.~\ref{fig:ngc3557_mom1}), coincident with an asymmetry in the PVD (Fig.~\ref{fig:ngc3557_PVD}). Furthermore, an increase of the gas velocity around the centre is visible at positive velocities. These features may result from various different phenomena (see Section~\ref{sec:discussion} for details). However, a velocity increase around the galaxy centre is plausibly associated with the Keplerian upturn arising from material orbiting around the central SMBH \citep[e.g.][]{Davis17}. Keplerian motions are expected in a potential dominated by a SMBH, and are visible in observations where the sphere of influence (SOI) of the SMBH is resolved and some emission arises from within the SOI. Spatially resolving the Keplerian region allows the SMBH mass to be constrained to high accuracy \citep[see e.g.][]{Onishi17,Davis17,Davis18}.

To explore this possibility, we first verified that the spatial resolution of the combined observations allows us to resolve the SOI of the SMBH. Assuming a central stellar velocity dispersion of $282$~km~s$^{-1}$ \citep{Brough07}, the SMBH mass would be $\log (M_{\rm SMBH})\simeq8.82$~M$_{\odot}$ (derived from the $M_{\rm SMBH}$ -- $\sigma_*$ relation of \citealp[][]{Tremaine02}). The resulting $R_{\rm SOI}$\footnote{$R_{\rm SOI} = \dfrac{G M_{\rm SMBH}}{\sigma_{\rm \ast}^{2}}$, where $G$ is the gravitational constant, $M_{\rm SMBH}$ is the SMBH mass and $\sigma_{\rm \ast}$ is the stellar velocity dispersion.} is $\approx60$~pc, only slightly smaller than the spatial resolution of the combined observations ($0.44''\approx90$~pc).  In order to take this feature into account and in the process to give an estimate of the SMBH mass at the centre of NGC\,3557, we model the gas velocity by using the approach adopted by \citet{Onishi17} and \citet{Davis17}.

\begin{figure*}
\centering
\begin{subfigure}[t]{0.3\textheight}
\centering
 \caption{}\label{fig:ngc3557_MGE_first}
\includegraphics[scale=0.4]{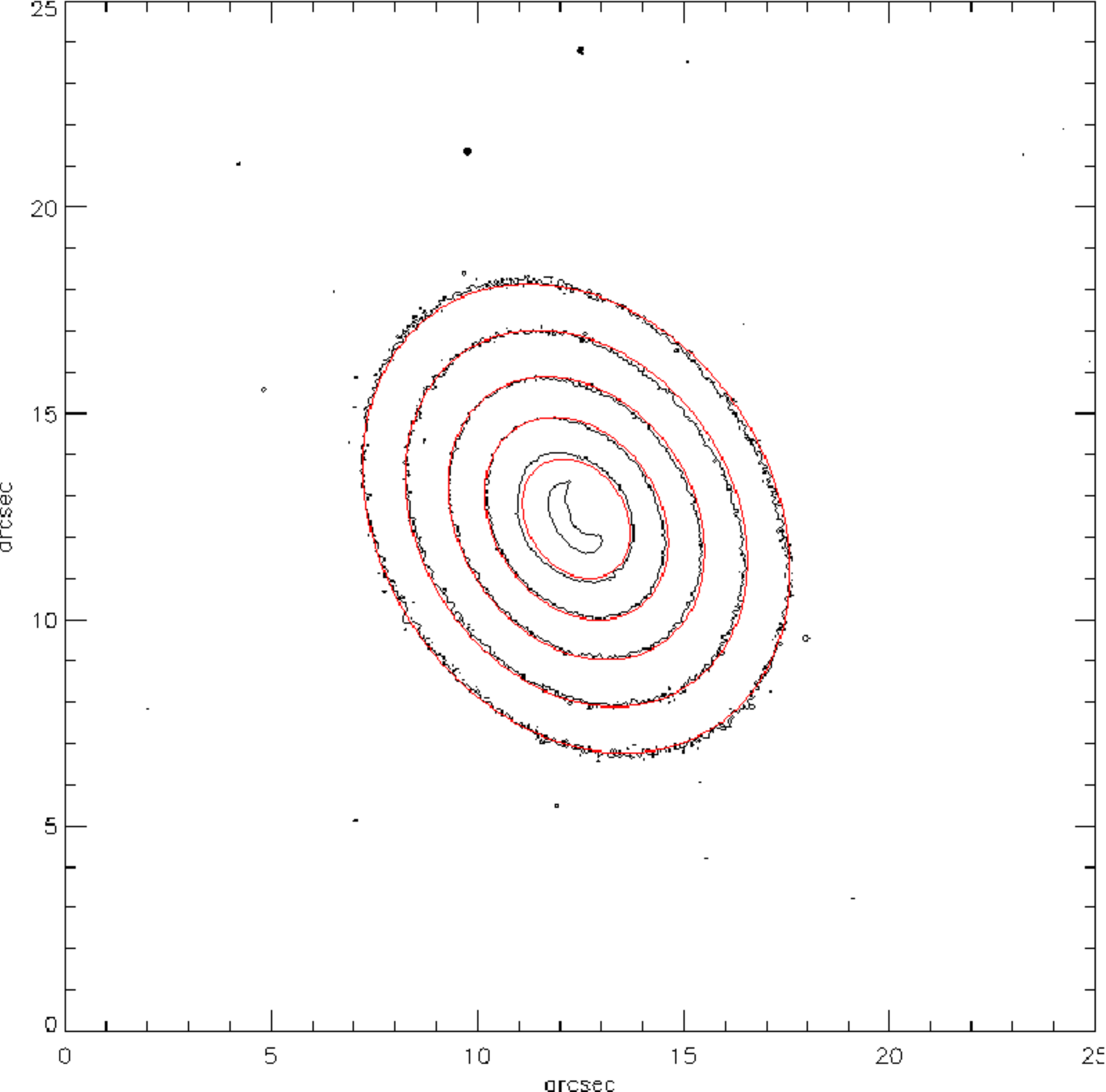}
\end{subfigure}
\hspace{10mm}
\begin{subfigure}[t]{0.3\textheight}
\centering
\caption{}\label{fig:ngc3557_MGE_second}
\includegraphics[scale=0.4]{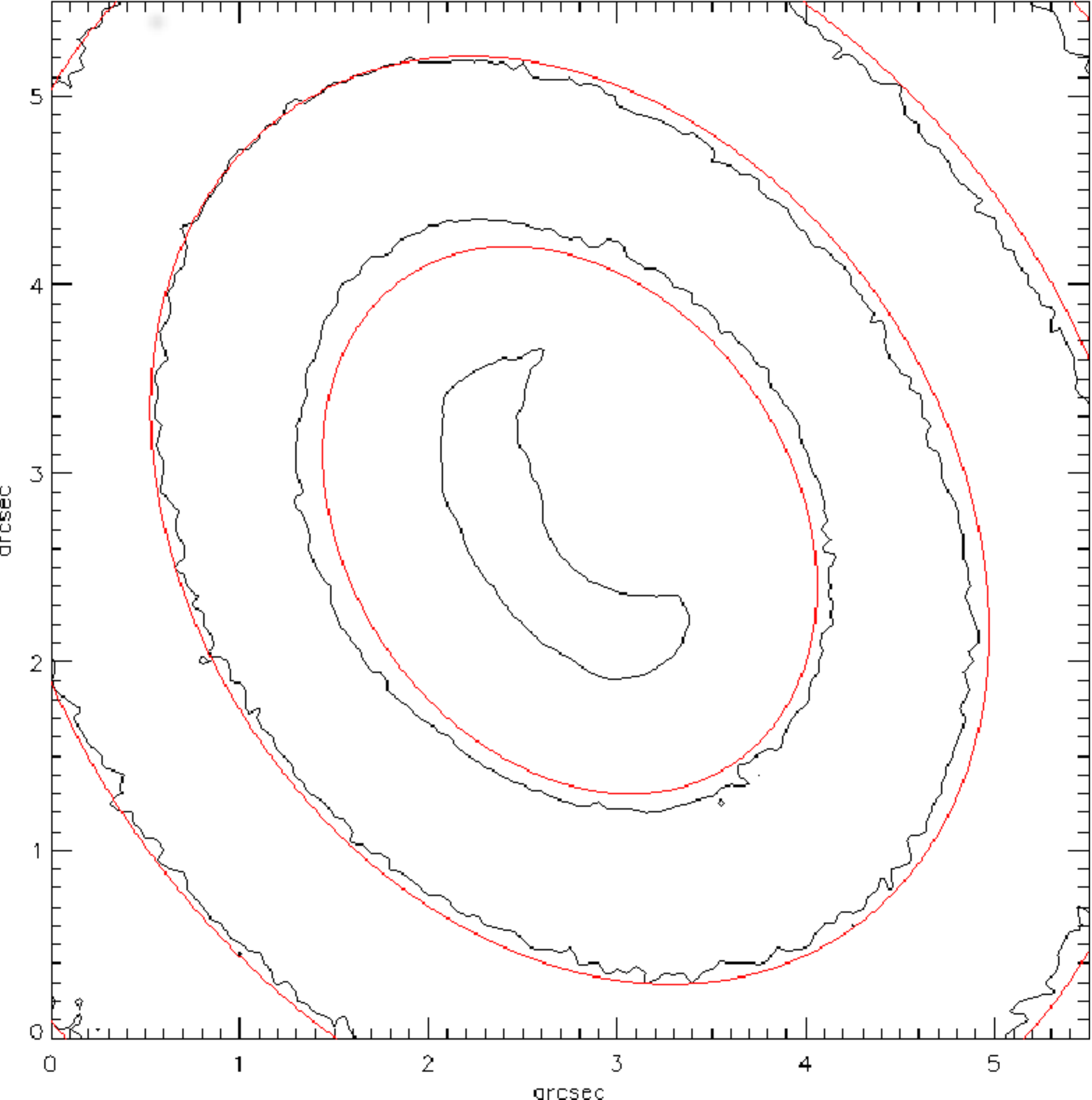}
\end{subfigure}
\caption{MGE model of NGC\,3557 (red contours), overlaid on a HST optical (F555W) image (black contours). Panel~\textbf{a} shows the whole galaxy, while panel~\textbf{b} shows a zoom in on the central region. The region masked due to dust is visible to the east of the galaxy nucleus in both panels. }\label{fig:NGC3557_MGE}
\end{figure*}

The mass budget in the central part of ETGs is dominated by their stars, so the kinematics of the molecular gas around the galaxy centre contains contributions from both the luminous stellar mass and the SMBH. To remove the contribution of visible matter and calculate the SMBH mass at the centre of NGC\,3557, we need to construct a mass model for the luminous component. This allows us to predict the circular velocity of the gas caused by the luminous matter, via the stellar mass-to-light ratio. The luminous matter distribution is parametrised using the multi-Gaussian expansion (MGE; \citealp{Emsellem94}) model adopted in the \textsc{MGE\_FIT\_SECTORS} package of \citet{Cappellari02}. In this technique, the three-dimensional stellar mass distribution of the galaxy is modelled by deprojecting a two-dimensional model of the observed stellar surface brightness and assuming a constant mass-to-light ratio ($M/L$). For our analysis we used the archival HST Wide-Field and Planetary Camera 2 (WFPC2) F555W image of NGC\,3557 (see Paper I, for details). Given the presence of a prominent dust ring seen at the centre of the galaxy, a longer wavelength high-resolution (e.g.\,HST) optical image would be needed to minimise dust extinction. However, such archival optical imaging is not available for NGC\,3557, so we attempt to mitigate the effect of dust obscuration by masking the dusty region. The best-fitting MGE model is shown in Figure~\ref{fig:NGC3557_MGE} and is tabulated in Table~\ref{tab:NGC3557_MGE_parameters}. The circular velocity curve arising from this mass model is then used to model the CO velocity curve in \textsc{KinMS}, with the SMBH treated as point mass. The inclination and position angles are fitted as single values throughout the disc. 
The molecular gas distribution is modelled as described in Section~\ref{sec:gas_distrib_kinem}. The best-fit model parameters of NGC\,3557 are listed in Table~\ref{tab:ngc3557}; a comparison between the observed and the modelled disc is shown in Figure~\ref{fig:NGC3557}.

The majority of the observed disc structures are reproduced well by our model, and accounting for the presence of a SMBH in the centre reproduces the velocity increase visible at positive velocities around the centre of the galaxy. We find a best fit $M_{\rm SMBH}=(7.10\pm0.02)\times10^{8}$~M$_{\odot}$ ($\log (M_{\rm SMBH})=8.85\pm0.02$~M$_{\odot}$), consistent with the expectation from the $M_{\rm SMBH}$ -- $\sigma_*$ relation \citep[e.g.][]{Tremaine02,Connell13}. Nevertheless, some significant residuals ($\approx 40$~km~s$^{-1}$) are visible in Figure~\ref{fig:ngc3557_mom1_res} at the centre of the gas distribution. The model predicts a symmetrical velocity increase at both positive and negative velocities, while only the former appears in the data. Thus, while the SMBH mass determined above seems reasonable, we cannot definitively claim that a massive dark object is the cause of the observed central velocity structures in NGC\,3557.

\begin{table}
\centering
\caption{MGE parameterisation of the NGC\,3557 light profile.}
\label{tab:NGC3557_MGE_parameters}
\begin{tabular}{c c c}
\hline
\multicolumn{1}{c}{ $\log_{10}I^{'}_{\rm j}$} &
\multicolumn{1}{c}{ $log_{10} \sigma_{\rm j}$ } &
\multicolumn{1}{c}{ $q^{'}_{j}$} \\
\multicolumn{1}{c}{ (L$_{\odot, i} pc^{-2}$) } &
\multicolumn{1}{c}{ ('') } &
\multicolumn{1}{c}{  } \\
\multicolumn{1}{c}{ (1) } &
\multicolumn{1}{c}{ (2) } &
\multicolumn{1}{c}{ (3) } \\
\hline
3.87  & 0.06 & 0.75 \\
 4.25  & 1.82 & 0.77 \\
 3.84  & 5.14 & 0.76 \\
\hline
\end{tabular}
\parbox[t]{8.5cm}{ \textit{Notes.} $-$ Columns: Intensity (1), width (2), and axis ratio (3) of each Gaussian component.}
\end{table}

\begin{table}
\centering
\caption{NGC\,3557 best-fit model parameters.}
\label{tab:ngc3557}
\begin{tabular}{l c c c}
\hline
\multicolumn{1}{c}{ ID } &
\multicolumn{1}{c}{ Parameter} &
\multicolumn{1}{c}{ Unit} &
\multicolumn{1}{c}{ Value } \\
\hline
(1) & Kinematic PA & (deg)  &  211$\pm$1.0  \\
 (2) & Inclination & (deg)  &  56$\pm1.0$ \\
(3) & $\sigma_{\rm gas}$ & (km~s$^{-1}$) &  10$\pm$1.0  \\
 (4) & v\textsubscript{offset} (v\textsubscript{cen}) & (km~s$^{-1}$)  &  15$\pm1.0$ (3014) \\
(5) & Stellar $M/L$ & ($M_{\odot}/L_{\odot}$)  &  5.4$\pm$0.1  \\
 (6)&  $\log (M_{\rm SMBH})$ & ($M_{\odot}$) & 8.85$\pm0.02$ \\
\hline
\end{tabular}
\parbox[t]{8.5cm}{ \textit{Notes.} $-$ Rows: (1) Kinematic position angle of the CO disc (i.e.\,the PA measured counterclockwise from North to the approaching side of the velocity field). (2) Inclination of the disc. (3)  Gas velocity dispersion. (4) Velocity offset of the kinematic centre (i.e.\,offset with respect to the expected and observed velocity zero-point; the expected zero-point corresponds to the redshifted frequency of the CO(2-1) transition listed in column (9) of Table~\ref{tab:summary}). The corresponding CO central velocity is reported in parentheses. (5) Stellar mass-to-light ratio. (6) Logarithm of the super-massive black hole mass.}
\end{table}

\subsection{NGC 3100}\label{sec:NGC3100}
The well-resolved CO(2-1) detection in NGC\,3100 shows a ring-like structure extending 1.6~kpc along the major axis, with a disruption to the North-West of the nucleus (see Paper I and Figure~\ref{fig:ngc3100_mom0}). The velocity field (Fig.~\ref{fig:ngc3100_mom1}) indicates that the kinematics of the gas are dominated by rotation, but with evident distortions in the rotation pattern. The iso-velocity contours are tilted and the major axis position angle changes moving from the edge to the centre of the ring, indicating the presence of a warp and/or non-circular motions. 

In our model, we initially adopted an exponential surface brightness profile instead of the Gaussian used above (an exponential form has been shown to be appropriate in many ETGs, e.g.\,\citealp{Davis13}). We also included an inner truncation (i.e.\,a hole). Such a simple profile does not fit the observed gas distribution adequately, leaving large residuals in the moment one maps. We find instead that the molecular gas distribution in NGC\,3100 is consistent with a two-armed spiral structure embedded in an exponential disc, with an inner truncation. Specifically, the disc profile is parametrised as

\[
   \Sigma_{\rm disc} \propto \left\{\begin{array}{lr}
	e^{-\dfrac{r}{R_{\rm disc}}} & \mbox{ ${\rm r} > {\rm R}_{\rm hole}$} \\
	0 & \mbox{ ${\rm r} \leq {\rm R}_{\rm hole}$}
  \end{array}\right.
\]

The spiral pattern is modelled as a uniform brightness distribution along loci whose X and Y positions from the phase centre are described by

\[
   \Sigma_{\rm spiral} \propto \left\{\begin{array}{l}
	x=[a~\cos\theta~e^{b\theta}, -a~\cos\theta~e^{b\theta}] \\
	y=[a~\sin\theta~e^{b\theta}, -a~\sin\theta~e^{b\theta}] \\
  \end{array}\right.
\]

The overall distribution has 5 free parameters in our \textsc{MCMC} code: the exponential disc scale length ($R_{\rm disc}$), the inner truncation radius ($R_{\rm hole}$), the constants of the parametric form of the logarithmic spiral function (a and b), and the angle of the spiral arms from the x-axis ($\theta$). 
This model provides a significantly better fit to the data, reproducing well also the main features shown in the observed integrated intensity (moment 0) map (Fig.~\ref{fig:ngc3100_mom0}). In particular, the regions of higher surface brightness at either side of the central hole are very well reproduced by the central spiral structure, as visible in Figure~\ref{fig:ngc3100_mom0_mod}.

We model the distortions in the rotation pattern by adding both a position angle and an inclination warp. We also assume that the disc is thin and the distribution axisymmetric. The velocity curve is parametrised following Equation~\ref{eq:arctan}. The gas is assumed to be in purely circular motion. Table~\ref{tab:ngc3100} summarises the best-fit ($\chi^{2}_{\rm red}=1.3$) model parameters. This model represents in general a good fit to the data (Fig.~\ref{fig:NGC3100}), reproducing most of the asymmetries in the rotation pattern. Nevertheless, the velocity field shows significant residuals, with peak amplitudes of $\approx\pm40$~km~s$^{-1}$. This suggests that the adopted model is not a complete representation of the gas kinematics and points towards the presence of non-circular motions.

\subsubsection{Non-circular motions}\label{sec:noncircular_motions}
\begin{figure*}
\begin{subfigure}[t]{0.3\textheight}
\centering
 \caption{Data moment zero}\label{fig:ngc3100_mom0}
\includegraphics[scale=0.3]{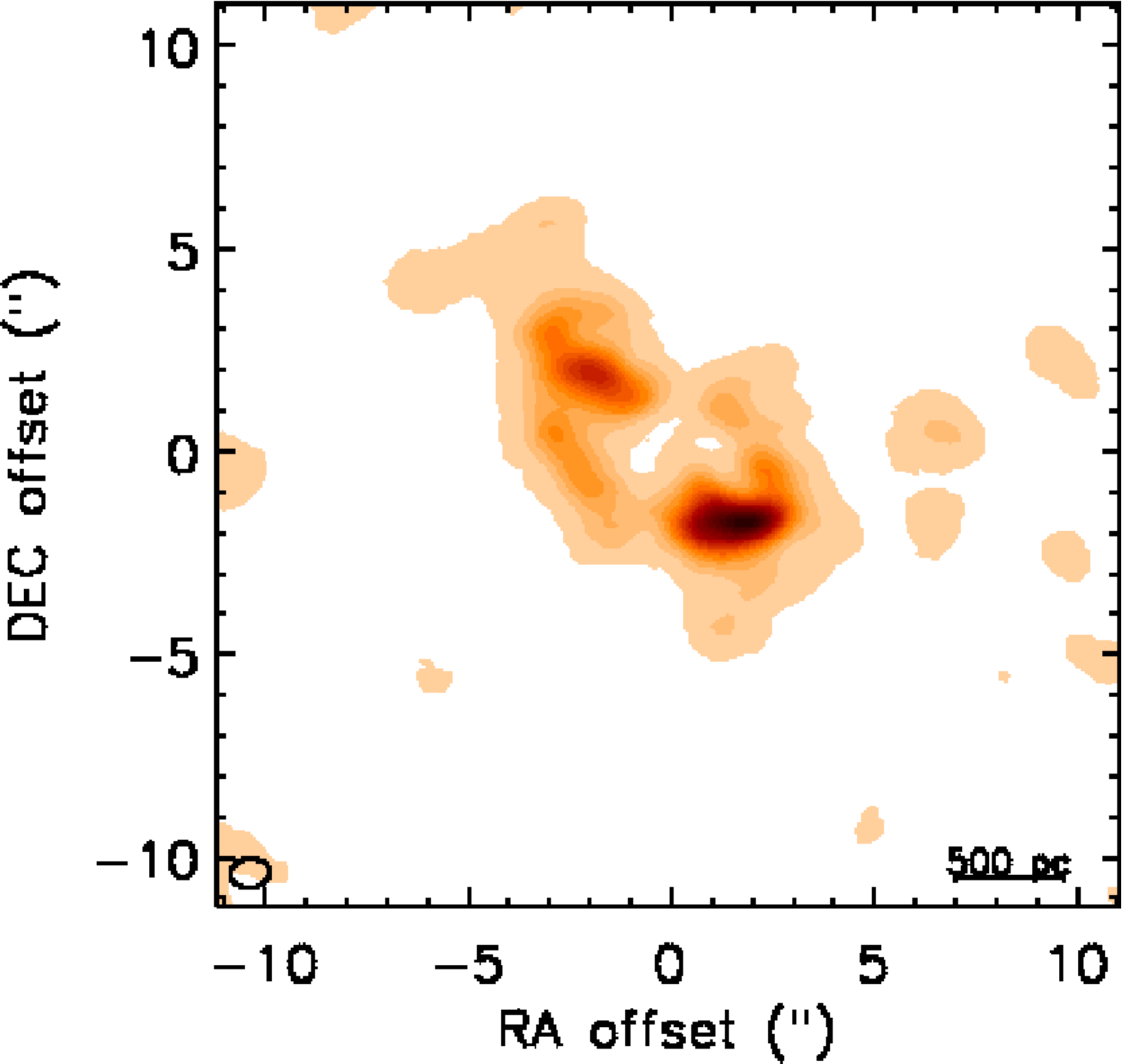}
\end{subfigure}
\hspace{10mm}
\begin{subfigure}[t]{0.3\textheight}
\centering
\caption{Model moment zero}\label{fig:ngc3100_mom0_mod}
\includegraphics[scale=0.3]{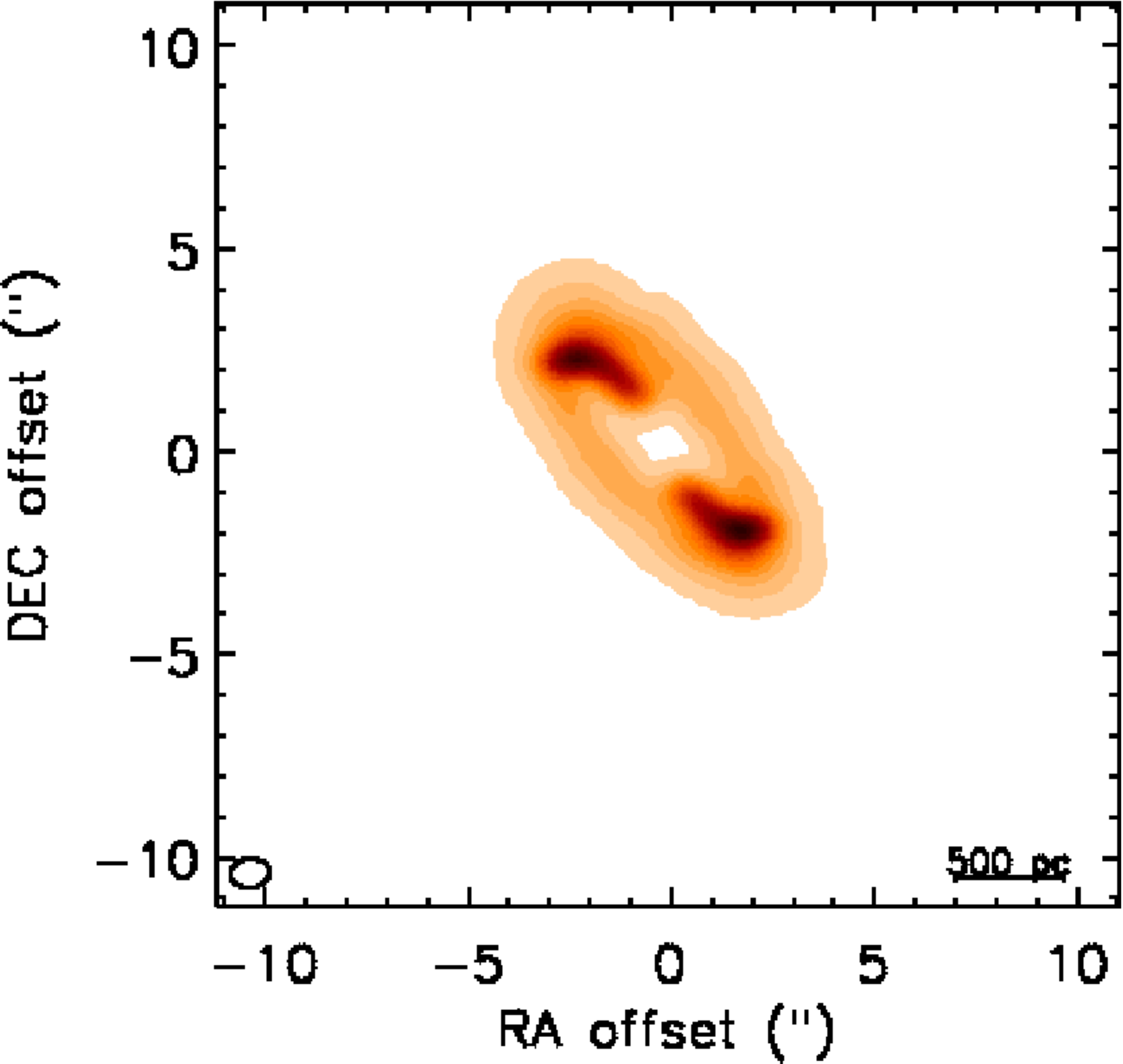}
\end{subfigure}

\medskip

\begin{subfigure}[t]{0.3\textheight}
\centering
 \caption{Data moment one}\label{fig:ngc3100_mom1}
\includegraphics[scale=0.385]{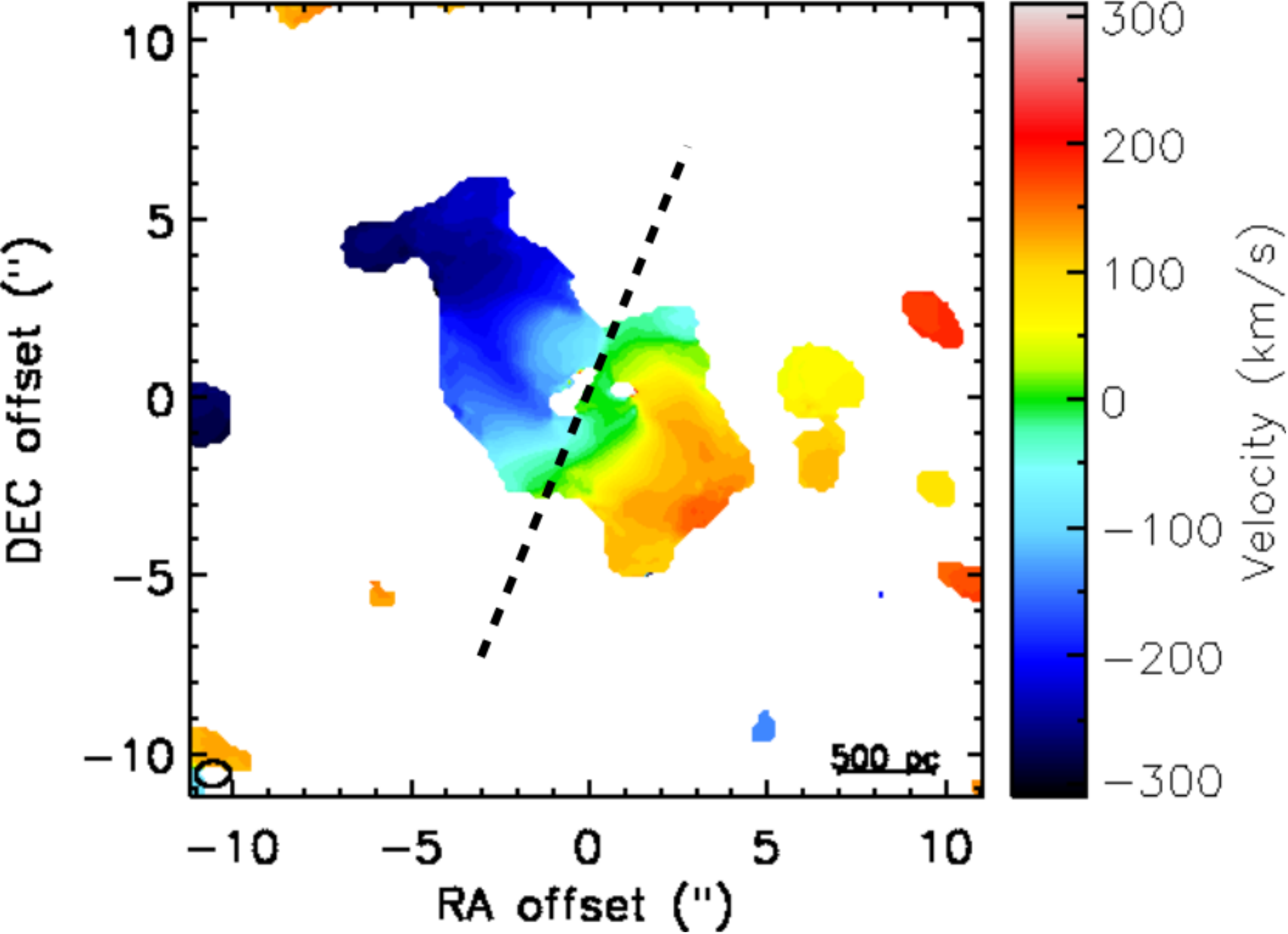}
\end{subfigure}
\hspace{12mm}
\begin{subfigure}[t]{0.3\textheight}
\centering
\caption{Model moment one}\label{fig:ngc3100_mom1_mod}
\includegraphics[scale=0.3]{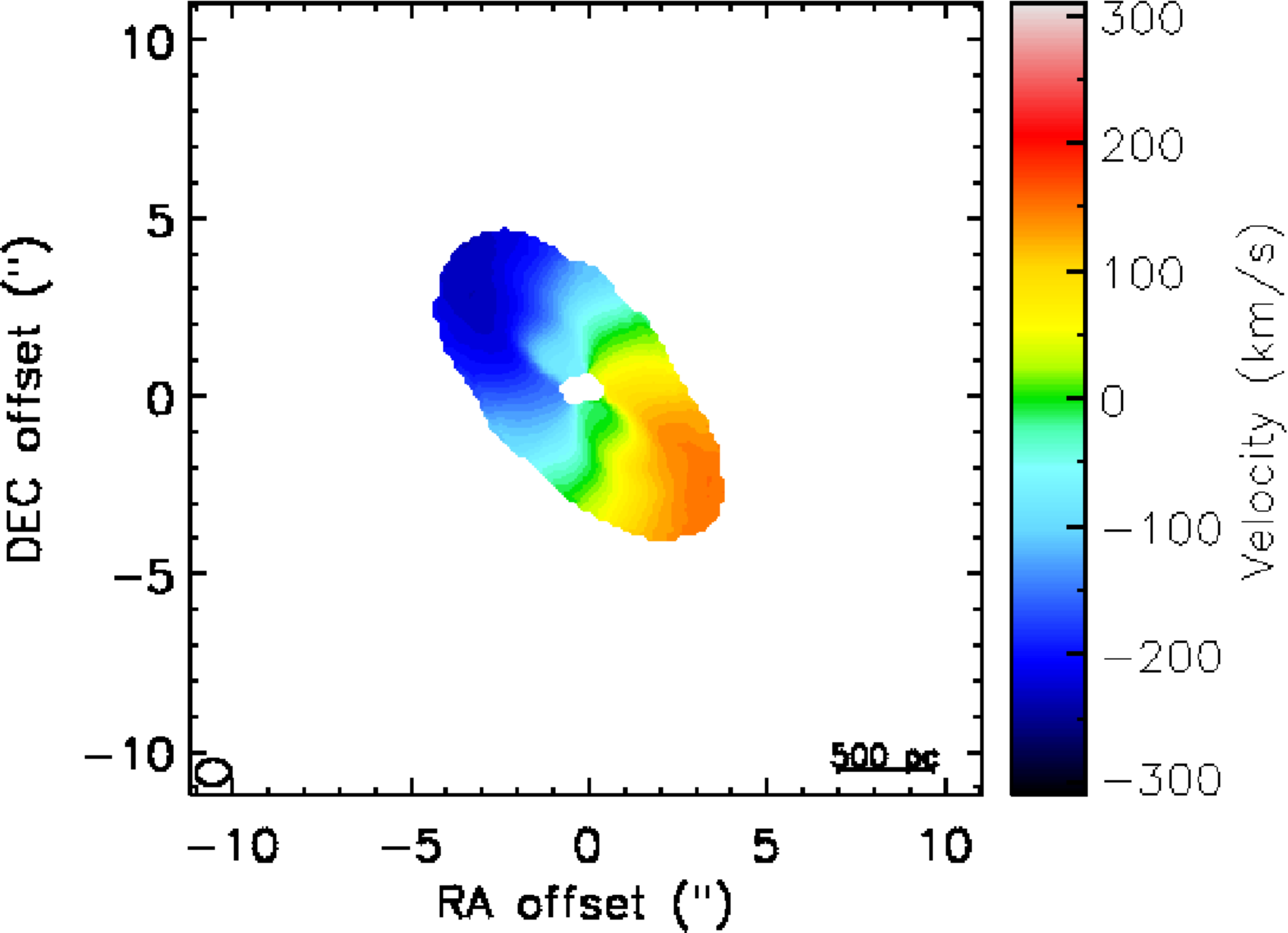}
\end{subfigure}

\medskip

\begin{subfigure}[t]{0.3\textheight}
\centering
\vspace{0pt}
\caption{Data - model moment one}\label{fig:ngc3100_mom1_res}
\includegraphics[scale=0.3]{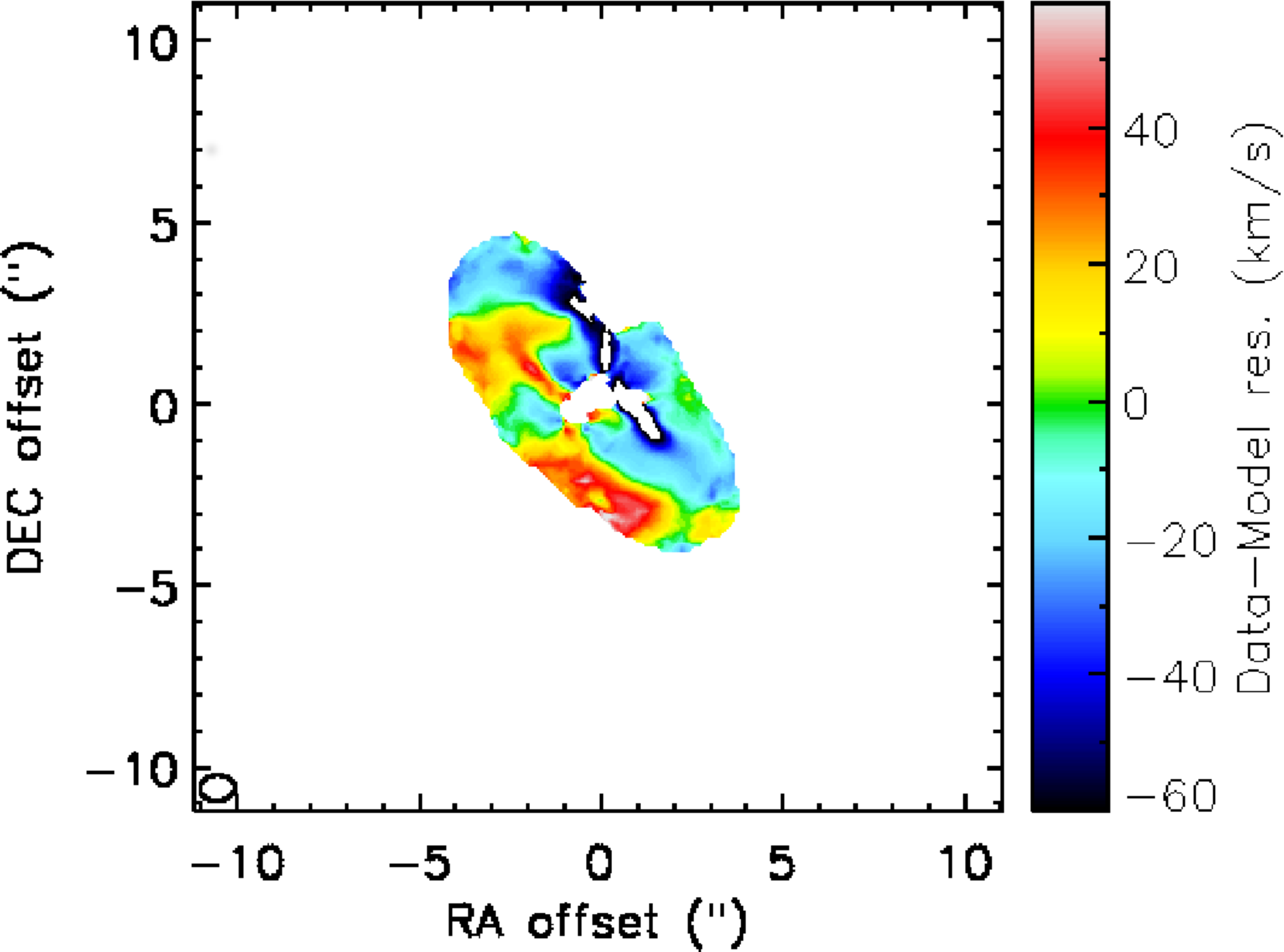}
\end{subfigure}
\hspace{10mm}
\begin{subfigure}[t]{0.3\textheight}
\centering
\vspace{0pt}
\caption{Position-velocity diagram}\label{fig:ngc3100_PVD}
\includegraphics[scale=0.3]{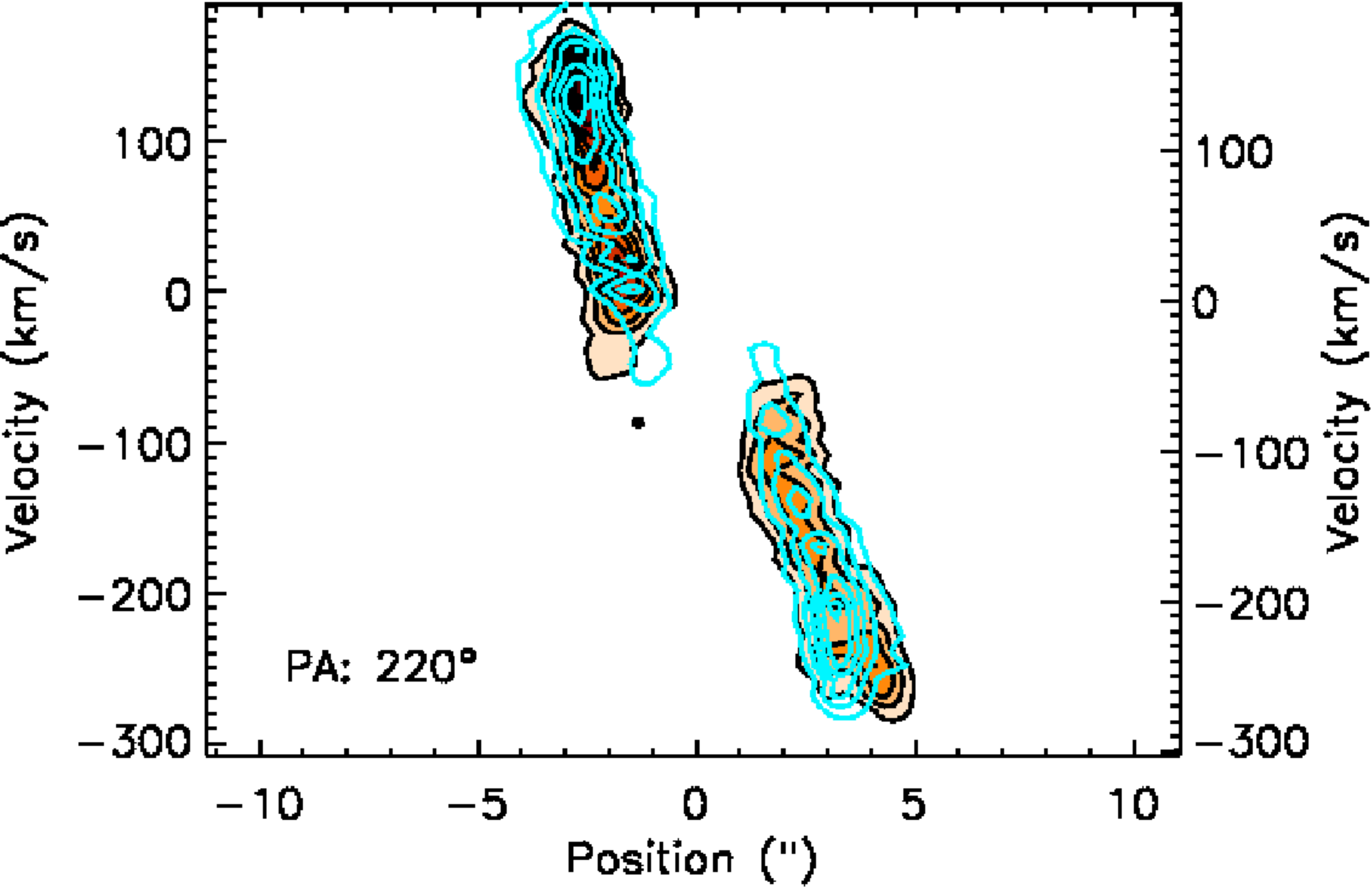}
\end{subfigure}
\caption{NGC\,3100 observed and simulated integrated intensity maps (moment 0, panels \textbf{a} and \textbf{b}), observed, mock, and residual mean velocity maps (panels \textbf{c}, \textbf{d}, and \textbf{e}) and PVD (panel \textbf{f}) as in Figure~\ref{fig:IC1531}, created using a data cube with a channel width of 10 km~s$^{-1}$.}\label{fig:NGC3100}
\end{figure*}

\begin{figure}
\centering
\includegraphics[scale=0.5]{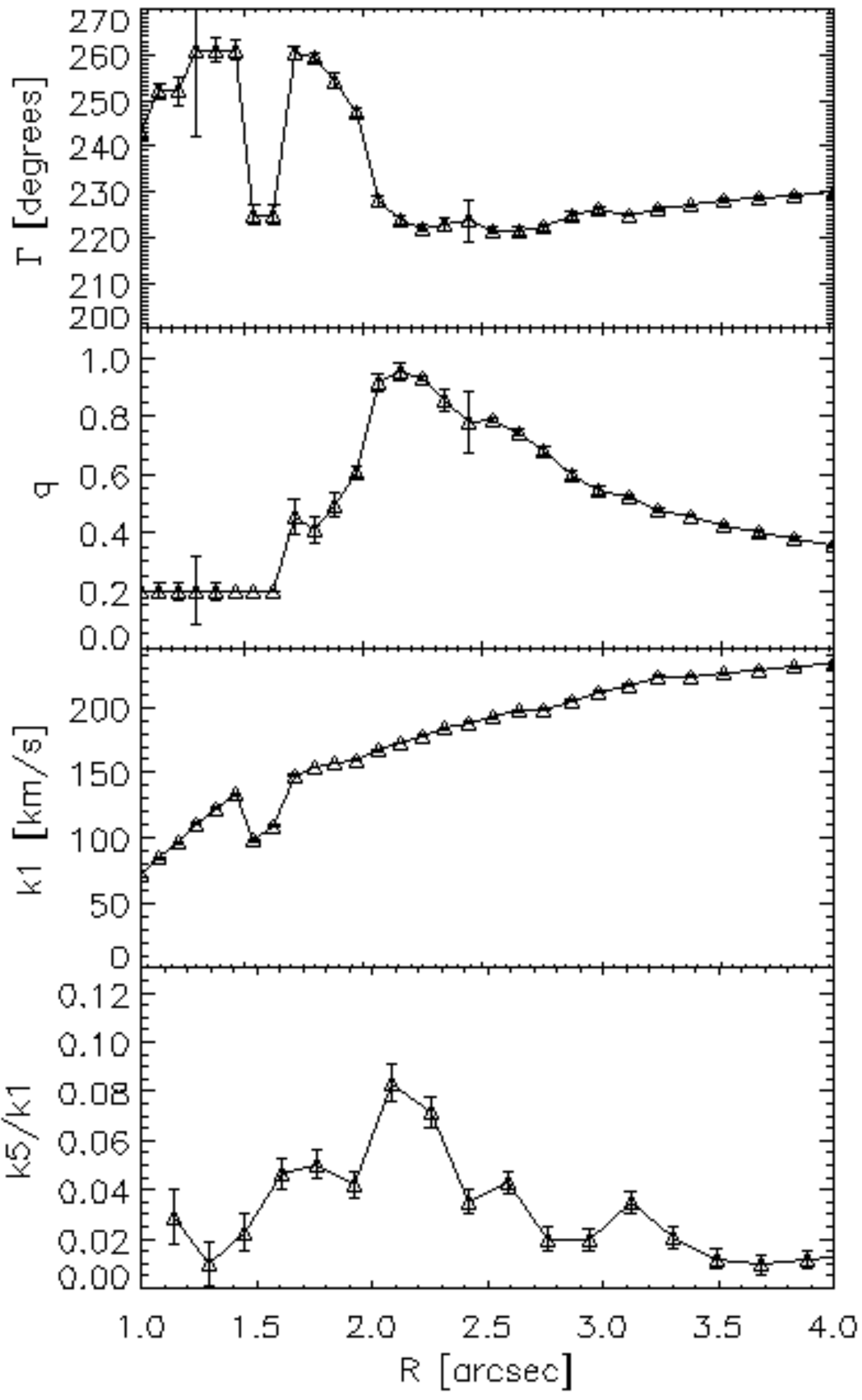}
\caption{Coefficients derived from the harmonic decomposition of the line-of-sight velocity field of NGC\,3100 (see Section~\ref{sec:noncircular_motions} for details). From the top to the bottom: kinematic PA; axial ratio of the best-fitting ellipses (directly related to the inclination in the approximation of a thin disc); harmonic term $k_{1}$, describing the rotation curve; $k_{5}$/$k_{1}$ ratio, characterising the level of non-circular motions with respect to the circular rotation. $k_{5}$ is the fifth-order term of the harmonic expansion and is used to trace the presence of non-circular motions. The x-axis indicates the distance from the nucleus along the disc major axis.}\label{fig:ngc3100_kinemetry}
\end{figure}

In order to characterise the nature of the residuals in the velocity field of NGC\,3100 (Fig.~\ref{fig:ngc3100_mom1_res}) and to check for the presence of non-circular motions (i.e.\,non-zero radial velocities) in the plane of the CO disc, we use the \textsc{Kinemetry} package \citep{Kraj06}. This method is a generalisation of surface photometry to the higher-order moments of the line-of-sight velocity distribution, determining the best-fit ellipses along which the profile of the moments can be extracted and analysed through harmonic expansion. Figure~\ref{fig:ngc3100_kinemetry} shows the kinemetric parameters measured for the observed velocity field of NGC\,3100. All the parameters are plotted versus the semi-major axis length of the fitted ellipses. According to the \textsc{Kinemetry} formalism, $\Gamma$ is the kinematic PA. The parameter $q$ is the flattening of the fitted ellipses, and can be interpreted as a measure of the opening angle of the iso-velocity contours. In the case of a thin disc (the assumption for NGC\,3100), the flattening is directly related to the inclination of the disc ($q=\cos i$). $k_{1}$ is the first-order term of the harmonic expansion and describes the amplitude of the bulk motions (i.e.\,the rotation curve). $k_{5}$ is the five-order term of the harmonic expansion, indicating deviations from simple circular rotation: this term is sensitive to the presence of separate kinematic components. In Figure~\ref{fig:ngc3100_kinemetry} the $k_{5}$ parameter is plotted in terms of a $k_{5}/k_{1}$ ratio, characterising the level of non-circular motions with respect to the circular rotation component. According to \citet{Kraj06}, a single kinematic component is defined as having slowly varying $\Gamma$ and $q$ radial profiles, with a nearly featureless velocity curve ($k_{1}$ term) and a $k_{5}/k_{1}$ ratio consistent with zero. Instead, separate kinematic components can be identified when observing an abrupt change either with $q>0.1$, or $\Gamma>10^{\circ}$, or with a double hump in $k_{1}$ with a local minimum in between. The transition between the different components is often emphasised also by a peak in $k_{5}/k_{1}$ (at the level of $\gtrsim0.1$), which thus serves as an additional signature for such a change.

The curves in Figure~\ref{fig:ngc3100_kinemetry} show that in the inner region ($r<1.5''$), the flattening is low, the circular velocity is rising and the $k_{5}/k_{1}$ ratio is mostly consistent with zero (i.e.\,$k_{5}\approx0$). The kinematic PA slightly changes in this region, suggesting the presence of a mild warp. The region between 1.5$''$ and 2$''$ is characterised by a sharp change of the PA ($>10^{\circ}$), likely indicating a kinematic twist. The presence of a kinematically distinct component in this region can be inferred by the slight drop of the circular velocity, accompanied by a rise of the $k_{5}/k_{1}$ ratio (i.e.\,$k_{5}>0$). The rising of the flattening parameter likely suggest that the two components (the inner and that within 1.5$''$-2$''$) are kinematically separate. Between 2$''$ and 2.5$''$ both the flattening and the $k_{5}/k_{1}$ ratio reach their maximum, whereas the PA has another abrupt change. Beyond 2.5$''$, q and $k_{5}/k_{1}$ decrease, with the latter becoming consistent with zero at the edge of the disc, and the PA remains almost constant. The velocity curve is rising and featureless from 2$''$ to the edge of the map. 

In summary, these complex features seem to indicate the presence of at least two separate kinematic components: an inner ($r<2.5''$) and an outer ($r>2.5''$) component. The harmonic decomposition also suggests changes in the kinematic PA ($\Gamma$) and inclination ($q$), consistent with the position angle and inclination warp we take into account in the disc modelling. In particular, at the best-fit, we find that the inclination ranges from $\approx80^{\circ}$ in the inner region to $\approx20^{\circ}$ towards the edges of the disc (Table~\ref{tab:ngc3100}), reasonably in agreement with the variation observed in $q$. Similarly, the position angle varies between $\approx$190$^{\circ}$ and $\approx$250$^{\circ}$ at the best \textsc{KinMS} fit. The presence of two separate kinematic components may also be compatible with the best-fit parametrisation of the CO distribution, consistent with an inner (r$<2.5''$) two-arms spiral structure embedded in an outer exponential disc. The $k_{5}/k_{1}$ ratio is non-zero, confirming that the radial motions are not negligible (10\% of the circular velocity at the peak) and this is likely to explain the 40~km~s$^{-1}$ residuals in the velocity map (Fig.~\ref{fig:ngc3100_mom1_res}). 

\begin{table}
\centering
\caption{NGC\,3100 best-fit model parameters.}
\label{tab:ngc3100}
\begin{tabular}{l l c c}
\hline
\multicolumn{1}{c}{ ID } &
\multicolumn{1}{c}{ Parameter} &
\multicolumn{1}{c}{ Unit} &
\multicolumn{1}{c}{ Value } \\
\hline
(1) &  Kinematic PA & (deg)  & 190$-$250  \\
(2) & Inclination & (deg)  &  20$-$80 \\
 (3) & v$_{\rm flat}$ & (km~s$^{-1}$) &  354$\pm$0.01    \\
(4) & r$_{\rm turn}$ & (arcsec)&  2.68$\pm$0.01 \\
 (5) & R$_{\rm disc}$ & (arcsec) &  3.36$\pm$0.01 \\
(6)  & R$_{\rm hole}$ & (arcsec) &  0.9$\pm0.02$     \\
 (7) & $\sigma_{\rm gas}$ & (km~s$^{-1}$) &  33.8$\pm$0.01  \\
 (8) & v$_{\rm offset}$ (v$_{\rm CO})$ & (km~s$^{-1}$)  &  91.5$\pm$0.01  \\
\hline
\end{tabular}
\parbox[t]{8.5cm}{ \textit{Notes.} $-$ Rows: (1) Range of the CO disc kinematic position angle (i.e.\,the PA measured counterclockwise from North to the approaching side of the velocity field). (2) Range of the disc inclination. (3) Asymptotic (or maximum) circular velocity in the flat part of the rotation curve. (4) Effective radius at which the rotation curve turns over. (5) Scale length of the exponential surface brightness profile. (6) Radius of the central surface brightness cut-off. (7) Gas velocity dispersion. (8) Velocity offset of the kinematic centre (i.e.\,offset with respect to the expected and observed velocity zero-point; the expected zero-point corresponds to the redshifted frequency of the CO(2-1) transition listed in column (9) of Table~\ref{tab:summary}). The corresponding central velocity is reported in parentheses.}
\end{table}

\section{Discussion}\label{sec:discussion}
\subsection{Kinematics of the molecular gas}
Our work clearly shows that sizeable amounts of cold gas are confined in discs in the inner (sub-)kpc of LERGs (Paper I). The 3D modelling of these discs reveals that the gas kinematics are dominated by circular rotation, but with evident distortions (both in the morphology and kinematics) that are mostly not fully described by simple axisymmetric models. Such disturbances can trace interesting kinematical properties. In general, they indicate the presence of unrelaxed substructures in the gas disc, possibly caused by the presence of either warps or non-circular motions, or a combination of both. Indeed, both phenomena produce the same kinematic features (e.g.\,tilted or s-shaped velocity iso-contours). Differentiating between the two through  disc modelling alone is not always straightforward, especially in discs where the gas distribution is only marginally resolved (as in IC\,1531, IC\,4296, NGC\,7075 and NGC\,3557). In cases like these, however, a combination of observational and theoretical views can provide useful hints to put constraints on the physical processes giving rise to the observed features.

 Lopsidedness, warps and distortions, like those seen in the rotation patterns of NGC\,612 and NGC\,3100 (Fig.~\ref{fig:ngc612_hole_mom1} and \ref{fig:ngc3100_mom1}, respectively) and in the velocity curve of NGC\,7075 (Fig.~\ref{fig:NGC7075_PVD}), are expected when the gas has been accreted externally (i.e.\,from interactions or minor merger events) and is still in the process of settling into the potential of the host galaxy \citep[e.g.][]{Lauer05}. This hypothesis will be extensively discussed in Section~\ref{sec:origin}. 
 
 Deviations from purely circular motions in gas discs can be induced by various other mechanisms. Among these are: bars in the stellar distribution;; inner spiral perturbations (often related to the presence of a large-scale bar) and interactions between the gas and the radio jets \citep[e.g.][]{Casasola11,Alatalo11,Combes13,Garcia14,Oosterloo17}. All of these processes can cause radial inflow or outflow.
 
 Jet-gas interactions produce strong asymmetries in both the gas distribution (e.g.\,disruptions in the gas distribution caused by fragmentation of the gas clouds) and kinematics (e.g.\,outflows), giving rise to features similar to those seen in the rotation pattern and velocity curve of IC\,1531 and NGC\,3557. Indeed, indications of the presence of a jet-cold gas interaction in NGC\,3557 have been recently claimed by \citet[][]{Vilaro19}. Interestingly, IC\,1531 and NGC\,3557 are the two sources in our sample showing a significant misalignment between the jet and the disc rotation axes (at least in projection; see Paper I), making a scenario in which the radio jets pierce the surrounding cold gas disc plausible. Furthermore, in both cases, the observed asymmetries seem to be located in the direction of the jet axis (outlined in Panel a of Figures~\ref{fig:IC1531} and \ref{fig:NGC3557}). We then speculate about a possible jet-gas interaction in these two sources, although higher-resolution data and/or other molecular gas transitions are needed to further explore thiss hypothesis.

Non-axisymmetric gravitational instabilities induced by the presence of stellar bars also leave clear signs in the morphology and kinematics of circumnuclear molecular gas distributions \citep[e.g.][]{Combes13}, compatible with those observed in NGC\,612 (Fig.~\ref{fig:NGC612} and Sect.~\ref{sec:individual_sources}). Cold gas discs in barred galaxies usually have nuclear gaps relying on bar-induced gravitational torques that force the gas outwards to the inner Lindblad resonance (ILR), or inwards towards the central SMBH \citep[e.g.][]{Combes01}. This process can also give rise to significant deviations from circular motions \citep[e.g][]{Combes01,Combes06,Randria15}, consistent with the presence of high residuals (with peak amplitudes of $\pm60$~km~s$^{-1}$) in the central part of the NGC\,612 CO disc (Fig.~\ref{fig:ngc612_hole_mom1_res}). Bars in edge-on galaxies also leave kinematical signatures in molecular gas PVDs, showing characteristic X-shapes (or simply sharp edges like those visible in Figure~\ref{fig:ngc612_hole_PVD}) that are indicative of the presence of two velocity components along the line of sight \citep[e.g.][]{Alatalo13}: an inner rapidly rising component associated with gas located within the ILR, and an outer slowly rising component associated with gas on nearly circular orbits beyond co-rotation. The disc in NGC\,612, however, is viewed almost edge-on (Table~\ref{tab:first_model}), hence projection effects (i.e.\,lines of sight passing through material at many different radii) prevent us from carrying out a detailed study of its complex kinematics. The study of the stellar light distribution (using high-resolution optical imaging) would be useful in this case: as bars also leave clear signatures also in the optical morphology and the stellar kinematics \citep[e.g.][]{Emsellem01}. However, the low spatial resolution optical images currently available for NGC\,612 (Paper I) prevents us from further investigating our hypothesis.

In summary, our analysis clearly shows that the observed molecular gas discs are not fully settled in a steady state into the stellar potential of the host galaxy. Although obvious observational limits do not allow us to draw strong conclusion in most of the cases, low-level dynamical disturbances can be identified in all the objects and may be associated with different underlying mechanisms (e.g.\,jet-gas interactions, stellar bars, external gas accretion, etc.\,). 

Disturbed CO discs are not commonly seen in local ETGs. In particular, significant differences are found between radio-loud (where the radio "loudness" is essentially associated with the presence of the radio jets) and radio-quiet objects. Interferometric observations of 40 nearby ETGs (mostly radio-quiet) imaged in CO(1-0) with the Combined Array for Research in Millimeter Astronomy (CARMA) in the framework of the ATLAS\textsuperscript{3D} survey show that 80\% of the discs are settled, both dynamically and morphologically \citep[i.e.\,they show regular rotation patterns and morphologies, without evident departures from symmetry or deviations from purely circular motions;][]{Alatalo13, Davis13}. Discs showing irregularities in their rotation pattern (such as s-shaped iso-velocity contours), warps, asymmetries and/or disruptions in their morphology (i.e.\,unsettled discs; 20\%) are objects classified as low-luminosity (or "radio-weak") AGN (on the basis of their 5~GHz radio emission; \citealp[][]{Nyland16}), interacting systems or objects that have acquired their gas externally. However, the CO discs in the CARMA sample are imaged at much lower resolution ($>4''$) than ours. Thus, although the proximity of these objects ($D_{\rm L}<46$~Mpc) allows to sample spatial scales comparable to that of our observations (a few hundred parsecs) in most of the cases, caution is needed in drawing conclusions from this comparison. High-resolution ($\leq$100~pc) ALMA observations of CO(2-1) discs in nearby ETGs have been presented by \citet{Boizelle17}. By analysing the kinematics of the five brightest CO discs (hosted by radio-quiet ETGs), they find that deviations from models assuming gas in purely circular motions are very small ($\lesssim10$~km~s$^{-1}$) and warping of the gas discs are of low magnitude in all the cases ($\lesssim10^{\circ}$), concluding that each CO distribution can be considered both dynamically and morphologically settled.

The emerging picture is that CO discs in LERGs (and, in general, in radio-loud objects) are significantly more disturbed than in radio-quiet objects. This provides additional clues to the potential connection between "jet-mode" AGN activity and the surrounding (sub-)kpc scale molecular gas discs and, more generally, to the interplay between the AGN and its host galaxy.

\subsection{NGC 3100: a case of fuelling?}\label{sec:ngc3100_discuss}
NGC\,3100 is certainly our most interesting case, particularly because it shows two separate kinematical components within the gas disc, the inner one associated with a two-armed spiral perturbation (Sect.~\ref{sec:NGC3100}).

Nuclear spirals are frequently observed in the cores of nearby ETGs \citep[e.g.][]{Lopes07}, particularly in those hosting LINER-like AGN \citep[e.g.][]{Martini03}, and are usually associated with shocks in nuclear gas discs \citep[e.g.][]{Martini99,Yuan04,Maciej04a,Maciej04b,Combes06,deVen10}. The gas is compressed along the spiral arms and the density enhanced. As a consequence, the gas surface brightness increases in the region of the spiral shock. This may explain the regions of higher surface brightness visible in the integrated CO intensity map of NGC\,3100 to either side of the central hole (Fig.~\ref{fig:ngc3100_mom0}), perfectly matching the location of the spiral arms in the best-fitting model (Fig.~\ref{fig:ngc3100_mom0_mod}).

Different mechanism have been suggested for the formation of nuclear spirals \citep[e.g.][]{Fathi11}. Non-axisymmetric potentials can cause a loss of the gas angular momentum due to torques and shocks, leading to the formation of inner spiral structures \citep[e.g][]{Maciej00,Emsellem01,Heller01,Shlosman01,Shlosman05,Maciej04a,Maciej04b,Fathi07}. As mentioned above, non-axisymmetric perturbations can be induced by large-scale stellar bars. The associated gravitational potential gives rise to a shock front along the bar edges, leading to an inward/outward flow of material that can form nuclear spiral structures  \citep[e.g.][]{Fathi06,Casasola11,Combes13}. However, the optical morphology and B-I colour map of NGC\,3100 (see Paper I) do not provide much evidence for the presence of a bar in this galaxy. We then consider it to be unlikely that a stellar bar is driving the inner spiral features. Another mechanism which may form nuclear spiral arms is shock formation induced by jet-gas interactions \citep[e.g][]{Veilleux05}. However, in this scenario, the location of the radio jets should be coincident with that of the spiral arms \citep[e.g.][]{Fathi11}, while in NGC\,3100 there is a significant misalignment between the two (see Paper I and Panel c of Figure~\ref{fig:NGC3100}). Non-axisymmetric perturbations can also occur if the molecular gas is accreted from outside the galactic system: in this case, the gas undergoes dynamical friction and then spirals towards the centre of the gravitational potential. In the central kpc-scales, while it partially continues to spiral in, the gas angular momentum may also spread it into a disc. Nuclear (dust and gas) spirals, larger-scale filaments and warped distributions, like those seen in NGC\,3100 (Fig.~\ref{fig:NGC3100} and Paper I), are typical tracers of this process \citep[e.g.][]{Malkan98,Lopes07}, indicating material that has been accreted externally and is not yet fully relaxed into the potential of the host galaxy. Indeed, NGC\,3100 is the best candidate among our sample sources for an external origin of the gas (see Section~\ref{sec:origin}), probably via interaction with the companion spiral galaxy NGC\,3095. We then favour external gas accretion as the mechanism giving rise to the spiral feature at the centre of the NGC\,3100 gas distribution.

Nuclear spirals are often invoked as a promising mechanism capable of transporting gas from kpc scales to the nucleus and feeding the central SMBH. This hypothesis been discussed many times, both from a theoretical and observational point of view \citep[e.g.][]{Wada02,Maciej04a,Maciej04b,Fathi06,Lopes07,Casasola08,deVen10,Hopkins10,Fathi11,Fathi13,Combes13}. One of the main problems linked to AGN fuelling is the removal of angular momentum from the gas. Spiral shocks and associated turbulence can induce streaming motions that cause the gas to lose or gain angular momentum and then to move inwards or outwards, respectively. The resulting radial motions can be a significant fraction of the underlying circular velocity \citep[e.g.][]{deVen10}. We then argue that the non-circular motions in the plane of the CO disc of NGC\,3100 (Sect.~\ref{sec:noncircular_motions}) may be associated with the inflow/outflow streaming motions induced by the spiral perturbation. In particular, the residuals in Figure~\ref{fig:ngc3100_mom1_res} exhibit mostly redshifted/blueshifted velocities to the eastern/western side of the CO ring, respectively. This implies gas inflow/outflow depending on the relative orientation of each side with respect to our line-of-sight. The B-I colour map of NGC\,3100 (Paper I) shows that the dust absorption is stronger at the eastern side of the ring, indicating that this is the near side. This suggest that the molecular gas is likely (at least partially) inflowing.

Given these considerations, it seems reasonable to conclude that the molecular gas at the centre of NGC\,3100 can feed the SMBH, fuelling the AGN. However, theoretical studies \citep[e.g.][]{Maciej04a,Maciej04b,Hopkins10,Hopkins11} indicate that spiral instabilities are capable of producing large accretion rates ($1 - 10$~M$_{\odot}$~yr$^{-1}$), comparable to those required to sustain the brightest radiatively efficient AGN. The typical inflow rates estimated for LERGs like NGC\,3100 indicate that, even when efficient mechanisms intervene to transport gas from kpc to $\sim$100~pc scales, other physical processes occur in the very inner regions of these objects, keeping the accretion rate low. Higher-resolution observations probing the gas at scales relevant for the accretion process ($\ll100$~pc) would be needed to examine the kinematics around the central SMBH and draw more solid conclusions.

Finally, we note that the anti-correlation between the jet synchrotron and CO emission northward to the nucleus, along with other observational evidence, make it plausible that there is a jet/gas interaction in this source (see Paper I for details). However, our modelling reveals no obvious kinematic signatures that can be attributed to such an interaction, likely indicating that, if present, this mechanism does not strongly affect the kinematics of the observed CO disc (at least at the current spatial resolution). A detailed analysis of the physics of the cold gas in NGC\,3100 through ratios of multiple molecular line transitions (recently observed with ALMA during Cycle 6) will allow us to further investigate this issue (Ruffa et al.\,, in preparation).

\subsection{Origin of the cold gas}\label{sec:origin}
The detection of a significant amount of molecular gas in the centre of LERGs, typically hosted by massive gas-poor ETGs, seems to point towards a recent regeneration of their gas reservoir \citep{Young14}. The origin of this gas, however, is still a subject of debate: it may be either internally generated or externally accreted. Disentangling these two scenarios may have important implications for our understanding of the fuelling mechanism of kinetic-mode AGN. 

In the first scenario, in situ cold gas formation may be the result of either stellar evolution (i.e.\,stellar mass loss) or cooling from the hot halo. Cold gas supply through stellar mass loss can always be present. In the most massive ETGs ($M_{*}>10^{11}$~M$_{\odot}$) evolved (primarily asymptotic giant branch) stars can inject gas and dust into the interstellar medium (ISM) at significant rates ($>3$~M$_{\odot}$~yr$^{-1}$, \citealp[e.g.][]{Jung01}). However, the detection rate of molecular gas in local ETGs seems to be independent of galaxy properties, such as stellar mass \citep[e.g.][]{Ocana10}, possibly suggesting that the majority of the gas in these systems has been accreted via different mechanisms. 

Large cold gas reservoirs can also be internally generated by cooling from the hot X-ray emitting gas phase, either smoothly (as predicted by, e.g.\, \citealp{Lagos14,Negri14,Lagos15}) or chaotically (as predicted in the framework of chaotic cold accretion models; see e.g.\, \citealp{Gaspari15,Gaspari17}). There is an established consensus that the hot halo provides the fuel source for LERGs, either directly or after cooling \citep[e.g.][]{Hardcastle07,Ching17,Hardcastle18,Gordon19}. This seems particularly plausible for LERGs in high density environments, where they are preferentially located \citep[e.g.][]{Best06,Hardcastle07,Best12,Sabater13,Ching17,Hardcastle18}. Indeed, a strong correlation is found between the intra-cluster medium (ICM) X-ray luminosity (a proxy of the environmental richness) and the 150~MHz radio luminosity of LERGs at $z<0.1$, providing evidence of a relationship between the ICM properties and the jet power of LERGs in groups and clusters and supporting a scenario in which the hot surrounding medium play a fundamental role in powering these systems \citep[e.g.][]{Ineson15}. Furthermore, kpc-scale filamentary or blob-like cold gas structures likely reminiscent of a hot gas cooling flow are often observed in massive nearby ETGs, most of which are LINER-like AGN associated with the brightest central galaxy (BCGs) of groups or clusters \citep[e.g.][]{David14,Werner14,Russell16,Temi18,Tremblay18,Nagai19}. The only comparable object in our sample is IC\,4296, located at the centre of a small cluster (Abell 3565). Kpc-scale blobs and filaments of cold gas are not detected within the field-of-view of our ALMA observations ($26''\approx6.7$~kpc at the source redshift). However, a recent analysis of the soft X-ray spectrum (0.2$-$0.8~keV) of IC\,4296 \citep[][]{Grossova19} finds evidence for gas cooling at a rate of $5.4$~M$_{\odot}$~yr$^{-1}$, concluding that the current rate of cooling is largely sufficient to produce the observed radio jets. It then seems plausible that the cold gas content of this object originates from radiative cooling of the hot halo. This would, however, require fast formation of dust grains in this cooled material in order to shield the molecular gas, and to create the obscuring structures seen in optical images (see Paper I). This is discussed further in Section \ref{dustformation}.

BCGs or central galaxies like IC\,4296, however, have a plentiful supply of gas from their surrounding large halos. All of the other LERGs in our current sample are in low density environments (i.e.\,isolated or in poor groups; see Paper I). The incidence of cooling from hot halos in isolated systems is not yet well understood. 
However, the fact that not all LERGs are in high density environments suggests that other factors can play a role in supplying their ISM reservoirs, such as galaxy-galaxy interactions \citep[][]{Sabater13} or minor merger events (although recently \citealp{Gordon19} argued against the importance of minor mergers in fuelling LERGs). Indeed, various observational studies conclude that the dominant mechanism for the gas supply in ETG hosts located in field or poor environments is external accretion \citep[e.g.][]{Kaviraj12,Davis15,Davis19}. Evidence for this can come from various indicators, such as morphological and kinematical disturbances in both the host galaxies and the central gas discs, the significant surplus of ISM compared to expectations from stellar mass loss, the wide range of gas-to-dust ratios and, in turn, metallicities with respect to that expected within internally generated ISM, the kinematic misalignments between gas and stars in these systems, or the absence of correlation between the gas/dust and galaxy properties  \citep[e.g.][]{Sarzi06,Annibali10,Davis11,Smith12,Kaviraj12,Alatalo13,Davis15,Duc15}. A detailed analysis of the cold gas origin based on extensive multi-wavelength information is beyond the scope of this paper. However, in the following we explore some of these indicators, in the attempt to put some constraints on the origin of the observed gas distributions.

\subsubsection{Stellar and gas kinematic (mis-)alignments}
Cold gas reservoirs that originate from internal processes are expected to co-rotate with the stars \citep[e.g.][]{Sarzi06}. For this reason, \citet{Davis11} proposed a criterion stating that misalignment angles $>30^{\circ}$ between the cold gas and stars rotation axes indicate an external origin of the gas. Indeed, cosmological simulations of isolated ETGs \citep[e.g.][]{Voort15} show that externally accreted cold gas discs can exhibit significant misalignments with respect to the stellar rotation axis (up to $180^{\circ}$). In two of our cases (NGC\,3100 and NGC\,7075) the CO and stellar velocity fields (observed with VIMOS/IFU; Warren et al.\,in prep.\,) show large misalignments ($>120^{\circ}$). This can, in principle, exclude a secular origin of the gas (i.e.\,from stellar mass loss), but not a hot gas cooling scenario. In fact, in cases where the spin of the halo is misaligned with respect to that of the stars, gas that is cooling  would not  necessarily be kinematically aligned with the stellar component, and  large discrepancies between the stellar and gaseous rotation axes ($>30^{\circ}$; \citealp[e.g.][]{Lagos15}) could be observed. While it is not clear how common such misaligned halos are in reality, the criterion of gas-star kinematic misalignment alone cannot establish the origin of the observed gas discs.  

\subsubsection{Distortions, warps and lopsidedness}
Gas discs that are dynamically relaxed and fully settled into the potential of the host galaxy are expected to show regular rotation patterns and compact disc- or ring-like morphologies, without evident departures from symmetry \citep[e.g.][]{Alatalo13}. Kinematical features such as warps or distortions, as well as filamentary or patchy distributions, are indicative of perturbations to a steady configuration and may suggest an external origin for the gas. Indeed, externally acquired gas takes time to settle down in the host galaxy environment, dynamically relax and co-rotate with stars. As in the settling sequence proposed by \citet[][]{Lauer05} and predicted by hydrodynamic simulations \citep[e.g.][]{Voort15}, during its evolution the gas streams towards the nucleus, experiencing torques due to the stellar potential. The stellar torques are stronger at the centre, where the gas therefore virialises faster than in the outskirts: this results in warps and asymmetries, like those visible in the NGC\,612 and NGC\,3100 discs (Figs.~\ref{fig:ngc612_hole_mom1} and \ref{fig:ngc3100_mom1}, respectively). The observed features, along with the gas/star kinematic misalignment, make it plausible that the cold gas in NGC\,3100 has been accreted externally and is still in the process of settling (see also Sect.~\ref{sec:ngc3100_discuss}). In NGC\,612, the molecular gas disc co-rotates with the stars. However, the disc is warped at its edges (Fig.~\ref{fig:ngc612_hole_mom1}) and H{\small \sc{I}} observations presented by \citet[][]{Emonts08} show the presence of a large-scale atomic gas disc (140~kpc along its major axis) connected with a bridge to the companion galaxy NGC\,619 (about 400~kpc away). As expected, the CO(2-1) disc occupies the central region of the atomic gas disc. It then seems reasonable to assume that NGC\,612 has also accreted its cold gas reservoir externally. We speculate that external gas accretion may also explain the kinematic misalignment between gas and stars and the asymmetries in the velocity curve of NGC\,7075 (Fig.~\ref{fig:NGC7075_PVD}), although in this case the resolution is too low to draw firm conclusions and other observational constraints are missing. 
 
\subsubsection{Relaxation timescales}\label{sec:trelax}
Assuming that the gas reservoirs of NGC\,612 and NGC\,3100 have been acquired externally, it is interesting to determine their position on the settling sequence. A rough estimate can be obtained by estimating the relaxation time ($t_{\rm relax}$; i.e.\,the time taken by a misaligned gas disc to settle into a stable configuration), assumed to be proportional to the dynamical time of the rotating disc at the transition between the unperturbed and the perturbed gas structures. Because the gas within the transition radius appears to be relaxed, whereas outside this radius it is clearly not settled yet, this time can give us clues on the age of the gas accretion event \citep[e.g.][]{Lauer05}. Models predict a wide range of $t_{\rm relax}$, varying from $10^{8}$~yr to the Hubble time depending on radius, with typical values for ETGs of $\approx5\times10^{8}$~yr \citep[e.g.][]{Voort15,Davis16}. Indeed, theoretical studies show that the relaxation process of unsettled gas discs in the potential of ETGs typically takes a few dynamical times ($t_{\rm dyn}$; \citealp{Tohline82,Lake83}). Specifically, it is found that the relaxation time is approximately $t_{\rm dyn}/\epsilon$, where $\epsilon$ is the eccentricity of the potential. For typical lenticular galaxies (like NGC\,612 and NGC\,3100) $\epsilon \approx 0.2$ \citep{Mendez08}. We therefore assume
\begin{eqnarray}
t_{\rm relax} \approx 5 \times t_{\rm dyn} = 5 \times \dfrac{2\pi R}{v_{\rm circ}}
\end{eqnarray}
where $R$ is the radius at the transition between the unperturbed disc and the perturbed gas structures, and $v_{\rm circ}$ is the corresponding deprojected rotational velocity. 

In NGC\,612 we assume $R=11''\approx7$~kpc and $V_{\rm rot}=400$~km~s$^{-1}$, estimating $t_{\rm relax}\simeq5.2\times10^{8}$~yr. This value is in reasonable agreement with theoretical studies for ETGs in poor environments \citep{Davis16}, and supports the idea that NGC\,612 is at an advanced stage of its settling sequence, with the central disc already relaxed within the host galaxy potential and co-rotating with stars, and its outskirts still in the process of settling.

The complex distribution and kinematics of the cold gas in NGC\,3100 make it hard to clearly identify a transition between relaxed and unsettled structures. However, knowing that nuclear rings are considered stable configurations in settling sequences \citep[e.g.][]{Lauer05}, we assume $R$ and $V_{\rm rot}$ at the transition between the inner ring and the warped structures at its edges (Fig.~\ref{fig:ngc3100_mom1}). Based on this assumption, $R=4''\approx1$~kpc, $V_{\rm rot}=200$~km~s$^{-1}$, and $t_{\rm relax} \simeq 9.5\times10^{7}$~yr. This short relaxation time suggests that NGC\,3100 is on an early stage of its settling process, and may indicate a very recent gas accretion event. The scenario of a recent gas injection is also consistent with the stars/gas kinematic misalignment and the distortions and large-scale structures observed both in the gas and dust distributions of NGC\,3100 (see also Paper I).

In both cases, however, we note that if the accretion of material is continuous, the relaxation process gets slower ($\approx80-100~t_{\rm dyn}$; \citealp[e.g.][]{Voort15,Davis16}), making our estimates lower limits for the true relaxation timescale.

\subsubsection{Dust and molecular gas co-spatiality}
\label{dustformation}
In Paper I we demonstrate that dust and molecular gas are co-spatial in all the four objects (NGC\,612, NGC\,3100, NGC\,3557 and IC\,4296) for which archival optical images were available, suggesting that they trace the same ISM. This observational evidence can have implications for the origin of the observed molecular gas distribution. 

Cold gas originating from the cooling of the hot halo is expected to be mostly dustless (with typical dust-to-gas ratios of $\approx10^{-5}$), at least initially. Indeed, dust grains, continuously injected within the hot ISM (T$\approx10^{7}$~K) by stellar ejecta, are destroyed by thermal sputtering on very short time scales ($\approx$10$^{7}-10^{8}$~yr; e.g.\,\citealp[][]{Mathews03a}). Dust can then grow in cold gas by accretion of condensable elements on to pre-existing grains, but this process usually takes place on $\approx$Gyr time-scales \citep[e.g.][]{Valentini15}. Dust-poor molecular gas structures originated from hot gas cooling has been observed in massive radio-loud ETGs at the centre of nearby cluster of galaxies (e.g.\,\citealp{Lim08,David14}). In these cases, the observed gas distributions do not correlate with the dust visible in optical images and/or associated colour (dust absorption) maps. 

This scenario implies that dusty cold gas distributions (like those observed in our four sample objects) can be still consistent with an origin from hot gas cooling, if it is "old" enough to let the dust grains form again within the cold gas clouds ($\approx$~Gyr timescales). Another possibility is that the dusty cold gas has a different origin, e.g.\, from external gas accretion \citep[e.g.][]{Davis11}. The observational constraints described above, along with the estimated relaxation timescale (see Section~\ref{sec:trelax}), strongly support the latter hypothesis for NGC\,612 and NGC\,3100. The cases of NGC\,3557 and IC\,4296 are less clear cut. In both objects, the cold gas co-rotates with stars and this is consistent either with an internal origin or with gas that has already settled into a stable configuration. In IC\,4296, there is also evidence for gas cooling from the hot halo \citep{Grossova19}, which could also give rise to the observed molecular reservoirs. In these cases, cooling of hot gas over a long period plausibly explains the observed dust cold gas reservoirs. Another option could be that the cold gas discs formed by a combination of internal and external formation mechanisms \citep[e.g.][]{Martini03}. Both scenarios, however, would result in similar observed gas features, becoming formally indistinguishable using the information currently available for these two objects.

\subsection{Gas disc stability}
Molecular gas distributions are typically unstable against gravitational collapse, causing a fragmentation into clouds and successive burst of star formation. Cold gas discs in early-type hosts, however, are generally found to be gravitationally stable, with a consequently lower nuclear star formation rate \citep[e.g.][]{Ho97}. The dynamical state of the observed gas discs can have important implications for the nuclear activity in LERGs. Indeed, the gravitational (in)stabilities of the gas discs could relate not only to the star formation activity, but also to the fuelling of the central SMBH: infall of molecular gas is expected if the gaseous disc is gravitationally unstable \citep{Wada92}.

The stability of a thin rotating gaseous disc against gravitational collapse can be assessed by calculating the Toomre parameter \citep{Toomre64}

\begin{eqnarray}
Q = \dfrac{\sigma_{\rm gas}\kappa}{\pi G \Sigma_{\rm gas}}
\end{eqnarray}
where $\sigma_{\rm gas}$ and $\Sigma_{\rm gas}$ are the velocity dispersion and surface density, respectively, of the molecular gas, $G$ is the gravitational constant, and $\kappa$ is the epicyclic frequency (i.e.\,the frequency at which a gas parcel oscillates radially along its circular orbit). The latter is calculated as  
\begin{eqnarray}
\kappa = \sqrt{4\Omega^{2}+R\dfrac{d\Omega^{2}}{dR}}
\end{eqnarray}
where $R$ is the radius of the gas distribution and $\Omega$ is the angular frequency ($\Omega=v_{\rm circ}/R$, with $v_{\rm circ}$ being the circular velocity). We estimate average values of the Toomre parameter along the CO discs by adopting the best-fit velocity dispersions listed in Table~\ref{tab:first_model} (assumed to be constant throughout the disc). The gas surface densities ($\Sigma_{\rm gas}$) are derived from those reported in Paper I (see also Table~5). $v_{\rm circ}$ is the bulk circular velocity. This is taken to be the velocity at which the gas rotation curve reaches its maximum and then flattens. We note that $\Sigma_{\rm gas}$ and $v_{\rm circ}$ need to be corrected for inclination, in order to take projection effects into account. We summarise the parameters and the corresponding values of Q in Table~\ref{tab:Toomre}.

Theoretically, the disc is expected to be unstable if Q$\lesssim$1, otherwise it is considered stable against gravitational collapse. In two cases (IC\,1531 and NGC\,612) Q is below unity. This is expected in NGC\,612, since it has been found to form stars at a rate of $\approx7$~M$_{\odot}$~yr$^{-1}$ \citep{Asabere16}. In IC\,1531 the low stability parameter could be mainly related to the very low value of the fitted velocity dispersion (Table~\ref{tab:first_model}). This, in principle, would indicate dynamically very cold gas \citep[e.g.][]{Davis17}, but such a low $\sigma_{\rm gas}$ value is not well constrained by the channel width of the IC\,1531 datacube (20~km~s$^{-1}$). In fact, in interferometric line observations with moderate S/N, the smallest velocity dispersion that is possible to measure is approximately $2\sqrt{2ln2}$ times smaller than the channel width, or $\approx9$~km~s$^{-1}$ in this case ($\approx$3 times the best-fit value reported in Table~\ref{tab:first_model}) . Similar considerations can be made for NGC\,7075. We derive Q$=$3.4 by adopting the best-fit velocity dispersion of 5.1~km~$^{-1}$, whereas the smallest velocity dispersion that is possible to constrain in this case is $\approx17$~km~s$^{-1}$ (channel width$=40$~km~$^{-1}$). In both cases, we then consider the Q values listed Table~\ref{tab:Toomre} as lower limits. In IC\,4296, Q is well above unity, and this is probably due to the very high best-fit velocity dispersion (Table~\ref{tab:first_model}). Theoretically, such high velocity dispersions indicate that the disc is stable against collapse. In this case, however, as the observed gas disc is only marginally resolved and viewed nearly edge-on, then the velocity dispersion is likely to be overestimated due to beam smearing, disc thickness and projection effects. Therefore, we consider the estimated Q as an upper limit in this case. 

Given the uncertainties in the best-fit gas velocity dispersions of IC\,1531, NGC\,7075 and IC\,4296, we make alternative estimates of Q assuming a canonical value of $\sigma_{\rm gas}=8$~km~s$^{-1}$ \citep[e.g.][]{Davis11,Voort18}, and find $Q\simeq0.4$, 5.5 and 3.6, respectively. These values show the same trend as those obtained from the best-fit $\sigma_{\rm gas}$, thus our conclusions would remain unchanged for either choice of dispersion. We should note, however, that in all the cases the low-level perturbations that are not taken into account in our modelling can modify the best-fit velocity dispersion used in our estimation, adding to the uncertainties in the derived Q values.

In the majority of the cases, the derived values for Q indicate that the discs are stable against gravitational fragmentation, consistent with what was found by \citet{Okuda05} for 3C\,31 (the prototypical LERG).
This supports a scenario in which the deep potentials of the massive host galaxies of LERGs (i.e.\,giant elliptical galaxies) stabilise the gas distributions against gravitational collapse. The mass transfer rate is likely to be low in gravitationally stable discs and this may be consistent with the low accretion rates in LERGs. Note, however, that Q values above unity (up to 12) have been found in nearby star-forming galaxies \citep{Romeo17} and that numerical simulations also show that gravitational instabilities cannot be completely excluded in regions where Q is slightly above unity \citep[e.g.][]{Li05}. We therefore conclude that the values in Table~\ref{tab:Toomre} do not allow us to put strong constraints on the dynamical state of the observed gas discs, although they provide useful indications.


\begin{table}
\centering
\caption{Toomre stability parameter and related quantities.}
\label{tab:Toomre}
\begin{tabular}{l c c c}
\hline
\multicolumn{1}{c}{ Target } &
\multicolumn{1}{c}{ $\Sigma_{\rm gas}$} & 
\multicolumn{1}{c}{ $v_{\rm circ}$ } &
\multicolumn{1}{c}{ $<Q>$ } \\
\multicolumn{1}{c}{ } &       
\multicolumn{1}{c}{  (M$_{\odot}$~pc$^{-2}$) } &
\multicolumn{1}{c}{  (km~s$^{-1}$) } &   
\multicolumn{1}{c}{   } \\ 
\multicolumn{1}{c}{   (1) } &   
\multicolumn{1}{c}{   (2) } &
\multicolumn{1}{c}{   (3) } &
\multicolumn{1}{c}{   (4) } \\

\hline
 IC 1531 & 3.2$\times10^{3}$  & 189  & >0.2 \\ 
  NGC 612  &  1.6$\times10^{3}$  & 303 & 0.1 \\
   NGC 3100  & 5.8$\times10^{2}$  & 169  & 2.5 \\
  NGC 3557  &  2.1$\times10^{3}$  &  268  & 1.5 \\
 IC 4296  & 1.0$\times10^{3}$ & 319 & $<$29  \\  
 NGC 7075 &  2.3$\times10^{3}$ & 417 & >3.4 \\
\hline
\end{tabular}
\parbox[t]{8.5cm}{ \textit{Notes.} $-$ Columns: (1) Target name. (2) Gas surface density corrected for the inclination ($\Sigma_{\rm gas}~cos(i)$, with $\Sigma_{\rm gas}$ being listed in Table~5 of Paper I). (3) Bulk circular velocity corrected for the inclination ($v_{\rm circ}/sin(i)$). (4) Estimated average Toomre stability parameters.}
\end{table}

\section{Summary}\label{sec:conclusion}
This is the second paper of a series aiming at investigating the multi-phase properties of a complete, volume- and flux-limited ($z<0.03$, S\textsubscript{2.7 GHz}$\leq0.25$~Jy) sample of eleven LERGs in the southern sky.

Here we present the study of the kinematics of six sample members detected in \co\  with ALMA. The results can be summarised as follows:
\begin{itemize}
\item By forward-modelling the data cubes, we find that the observed discs are described reasonably well by simple axisymmetric models assuming gas in purely circular motion, indicating that the bulk of the molecular gas is in ordered rotation. Nevertheless, low-level deviations (both in the morphology and kinematics) are found in all cases (even in those which are poorly resolved), suggesting that the gas is not completely relaxed into the potential of the host galaxy and pointing towards the presence of non-circular motions in some cases. 
\item In IC\,1531 and NGC\,3557, the asymmetries observed in the CO rotation pattern and velocity curve lead us to speculate about the possibility of a jet-cold gas interaction. Indeed, indications about the presence of a jet/ISM interaction in NGC\,3557 have been recently proposed by other authors. Moreover, in both cases, the observed asymmetries seem to be located along the jet axis (at least in projection), reinforcing our hypothesis.
\item The velocity curve of NGC\,3557 shows an increase of the gas velocity around the centre, which we associate with the Keplerian upturn arising from material orbiting around the SMBH. This allows us to estimate a SMBH mass of $(7.10\pm0.02)\times10^{8}$~M$_{\odot}$ at the centre of this source, in agreement with expectations from the M$_{\rm SMBH}-\sigma_{*}$ relation. 
\item We find that the CO in NGC\,612 is distributed in a disc-like structure with a central gap.  The central gas deficiency, along with the sharp edges in the velocity curve, lead us to suppose the presence of a stellar bar in this object. This is also consistent with the high residuals ($\pm60$~km~s$^{-1}$) visible in the velocity map at the centre of the CO disc, indicating the presence of non-circular motions, expected in presence of non-axisymmetric potential due to a bar.   
\item NGC\,3100 is the most interesting of our CO detections. The best-fitting gas distribution is consistent with a two-armed spiral embedded in an exponential disc. The distortions in the rotation pattern are best described by the combination of warps in both position angle and inclination, although high residuals ($\pm40$~km~s$^{-1}$) in the velocity maps suggest the presence of non-circular motions in the plane of the CO disc. A harmonic decomposition of the line-of-sight velocities indicates that these high residuals are consistent with the presence of non-negligible radial motions ($\leq$10\% of the rotational velocity), that we link with the streaming motions associated to the spiral perturbations in the inner CO disc. The location of the disc along the line of sight suggests that the dominant radial motions are inwards towards the nucleus, supporting a scenario in which the observed cold gas is contributing to the fuelling of the AGN activity.
\item A number of observational and theoretical constraints allow us to conclude that an external origin of the cold gas (i.e.\,via interaction with the companion galaxies) is strongly favoured in NGC\,612 and NGC\,3100. Based on simple arguments, we roughly estimate their relaxation times (i.e.\,the times taken by externally acquired gas to relax dynamically and settle into a stable configuration) to be $\simeq5.2\times10^{8}$~yr and $\simeq9.5\times10^{7}$~yr, respectively, indicating that they are respectively in advanced and early stages of their settling sequences. An external gas origin is also plausible for NGC\,7075, but our results are not definitive. The origin of the cold gas in NGC\,3557 and IC\,4296 is unclear, although it is plausible that it originate from a combination of internal (i.e.\,hot gas cooling) and external (i.e.\,interaction or minor merger) processes. Given that all but one of the sources in our sample inhabit poor environments (whereas LERGs are preferentially found at the centres of groups and clusters), an external origin of the gas may have important implications for the powering scenario of LERGs in low density environments.
\item By calculating the average Toomre stability parameter of the CO discs, we find that in one case (NGC\,612) the disc can be certainly considered gravitationally unstable ($Q<1$). This implies that the gas can collapse to form stars (as found by other authors in this object), but can also relate to the fuelling of the central SMBH. In all the other objects, the $Q$ values are $>1$, suggesting that the disc are stable against gravitational collapse. This is consistent with the low accretion rate in LERGs, although various studies demonstrate that Q values similar to that we derive do not allow to put strong constraints on the dynamical state of the gas disc. 
\end{itemize} 


\section*{Acknowledgements}
We thank the anonymous referee for useful comments. This work was partially supported by the italian space agency (ASI) through the grant "Attivit\'{a} di Studio per la comunit\'{a} scientifica di astrofisica delle alte energie e fisica astroparticellare"
 (Accordo Attuativo ASI-INAF n.2017-14-H.0). TAD acknowledges support from a Science and Technology Facilities Council UK Ernest Rutherford Fellowship. IP acknowledges support from INAF under the PRIN SKA project FORECaST. This paper makes use of the following ALMA data: ADS/JAO.ALMA\#[2015.1.01572.S] and ADS/JAO.ALMA\#[2015.1.00878.7]. ALMA is a partnership of ESO (representing its member states), NSF (USA) and NINS (Japan), together with NRC (Canada), NSC and ASIAA (Taiwan), and KASI (Republic of Korea), in cooperation with the Republic of Chile. The Joint ALMA Observatory is operated by ESO, AUI/NRAO and NAOJ. The National Radio Astronomy Observatory is a facility of the National Science Foundation operated under cooperative agreement by Associated Universities, Inc.
 This paper has also made use of the NASA/IPAC Extragalactic Database (NED) which is operated by the Jet Propulsion Laboratory, California Institute of Technology under contract with NASA. This research used the facilities of the Canadian Astronomy Data Centre operated by the National Research Council of Canada with the support of the Canadian Space Agency.



\bibliographystyle{mnras}
\bibliography{mybibliography} 




\bsp	
\label{lastpage}
\end{document}